\documentclass[aps,prb,twocolumn,floats,pdflatex]{revtex4-1}
\usepackage{amssymb}
\usepackage{amsbsy}
\usepackage{amsmath}
\usepackage{amsfonts}
\usepackage{graphicx}
\usepackage{amssymb}
\usepackage{upgreek}
\usepackage{dsfont}
\usepackage{bm}
\usepackage{color}
\usepackage{hyperref}

\begin{document}

\title{Complete optical valley polarization in Weyl semimetals in strong magnetic fields}

\author{Simon Bertrand}
\author{Jean-Michel Parent}
\author{Ren\'e C\^ot\'e}
\author{Ion Garate}
\affiliation{Institut Quantique,  Regroupement Qu\'eb\'ecois sur les Mat\'eriaux de Pointe et D\'{e}partement de Physique, Universit\'{e} de Sherbrooke, Sherbrooke, Qu\'{e}bec, Canada J1K 2R1}

\date{\today}

\begin{abstract}
We present a theory of an optically induced valley polarization in an interacting, time-reversal symmetric Weyl semimetal placed under strong magnetic fields.
Because the application of a magnetic field reduces the symmetry of the crystal, the optical absorption intensity differs at Weyl nodes that were equivalent by symmetry at zero field. 
At strong magnetic field, the difference in the absorption intensity reaches 100\% for a sizeable frequency interval of the incident light. 
This complete valley polarization originates from interband transitions involving the chiral Landau level, and can be controlled 
by changing the directions of the magnetic field and the light propagation.
We identify the splitting of $0\to 1$ or $-1\to 0$ inter Landau level transitions as an observable signature of the complete valley polarization, and discuss its manifestation in the TaAs family of materials.
\end{abstract}
\maketitle

\section{Introduction}
\label{sec:intro}

One striking property of Weyl semimetals (WSM) in strong magnetic fields is the existence of chiral Landau levels.\cite{Nielsen1983, Armitage2018}
A chiral Landau level is special in that it is topologically protected, it intersects the Fermi energy only once, and electrons therein propagate unidirectionally along the magnetic field.

Chiral Landau levels are responsible for a number of anomalous transport and optical properties.
These include negative magnetoresistance,\cite{SonSpivak2013, Zhang2016, Arnold2016, Huang2015, Wang2016, Shekhar2015, Du2016} unusual collective modes,\cite{Panfilov2014, Song2016, RinkelPLS2017, RinkelPLS2019, Long2018} strong photocurrents,\cite{Golub2018} and robust transport of bosonic excitations in metamaterials.\cite{Jia2019}

Recent studies\cite{Jiang2018, Yuan2018} have reported on the first direct detection of chiral Landau levels using reflectivity measurements.
These experiments have been carried out in simple configurations, with the wave vector ${\bf q}$ of the incident light and the external (quantizing) magnetic field ${\bf B}$ parallel to the $c$ axis of the crystal.
The results of the measurements have been interpreted on the basis of simple model Hamiltonians.
Though most of the data fits well to the theory, the authors of Ref. [\onlinecite{Jiang2018}] observe an unusual splitting in certain inter Landau level transitions.

Partly motivated by these experiments, we develop a more complete theory of the interband magneto-optical absorption in WSM, with an eye on new effects that may emerge from chiral Landau levels.
Our work goes beyond earlier theoretical studies\cite{Ashby2013, ShaoYang2016, Sun2017} by exploring general orientations of ${\bf B}$ and ${\bf q}$, by analyzing the symmetry-reducing effect of the magnetic field and the incident light, and by incorporating electron-electron interactions.

The main new result of this work is the prediction of a complete valley polarization that is induced optically in WSM subjected to strong magnetic fields.
By ``complete'', it is meant that
the optical absorption takes place only in some of the symmetry-equivalent Weyl nodes. 
This effect, which originates exclusively from interband transitions involving the chiral Landau level, requires either a tilt of the Weyl cones or nonlinear terms in the energy spectrum.

The valley polarization can  be controlled by changing the directions of ${\bf B}$ and ${\bf q}$.
In particular, when the magnetic field points along a low symmetry direction and the incident light has the appropriate frequency, the entire optical absorption can be concentrated in the vicinity of a {\em single} Weyl node.
The main experimental consequence of the valley polarization is a splitting of the inter Landau level transitions involving chiral Landau levels, a feature reminiscent of that observed in Ref. [\onlinecite{Jiang2018}].

The rest of the paper is organized as follows.
Section \ref{sec:mod} presents a minimal model for an interacting Weyl semimetal with broken inversion symmetry.
Section \ref{sec:res} shows the results for the optical absorption in the minimal model.
Section \ref{sec:Disc} extrapolates the results from Sec.~\ref{sec:res} to real Weyl semimetals belonging to the TaAs family, and discusses the connection with recent experiments.
Section~\ref{sec:conc} summarizes the main ideas and provides an outlook.
The Appendices collect technical details, mainly pertaining to the calculation of the optical conductivity within the generalized random-phase approximation.

\section{Minimal model}\label{sec:mod}

\subsection{Low-energy Hamiltonian}\label{subsec:model}

\begin{figure}[t]
\includegraphics[width=0.45 \textwidth]{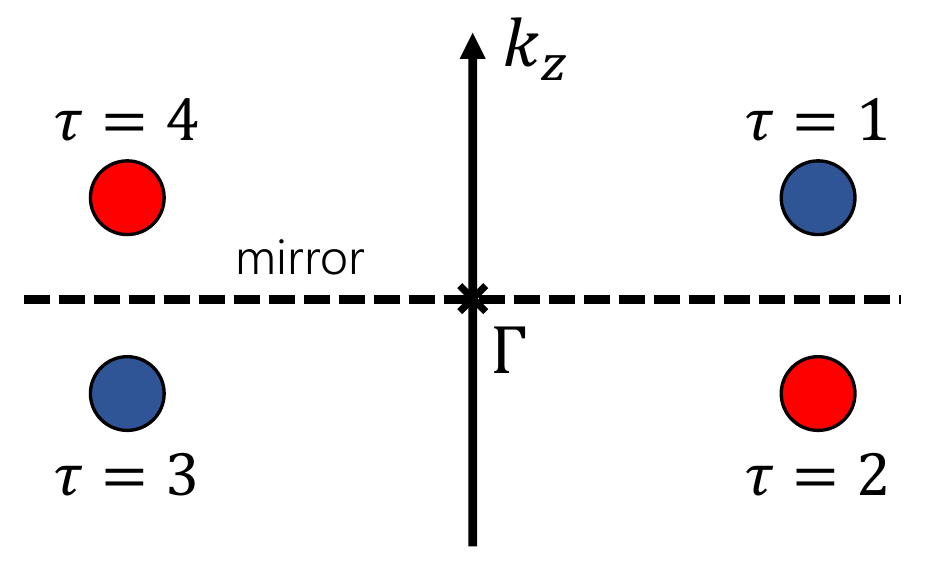}
\caption{
Schematic representation of a minimal four-node Weyl semimetal with time-reversal symmetry and a mirror symmetry.
The blue (red) dots represent nodes of chirality $+1$ ($-1$). The dashed line indicates a mirror plane and the $\Gamma$ point is a time-reversal invariant point. The nodes $\tau=1$ and $\tau=3$ are related by time-reversal symmetry, whereas nodes $\tau=1$ and $\tau=2$ are related by the mirror.}
\label{fig:nodes}
\end{figure}

We adopt a toy model for a WSM with time-reversal symmetry and broken spatial inversion symmetry.
This model contains four tilted Weyl nodes, two for each chirality.
We assume that the pairs of nodes of opposite chirality are related to one another by a mirror plane perpendicular to the $z$ direction (see Fig.~\ref{fig:nodes}).
Because of the mirror and time-reversal symmetries, the four nodes are symmetry-equivalent in the absence of a magnetic field.
We do not assume the presence of any rotation axis relating the different nodes.

The noninteracting low-energy Hamiltonian of such WSM at zero field can be expressed as
\begin{equation}
  \label{eq:H0}
  {\cal H}_0 = \sum_{{\bf k} \sigma \sigma' \tau} h_{\sigma\sigma' \tau}({\bf k}) c_{{\bf k}\sigma\tau}^\dagger c_{{\bf k}\sigma'\tau},
\end{equation}
where $c^\dagger$ and $c$ are the fermion creation and annihilation operators, ${\bf k}$ is the momentum measured with respect to a node,
\begin{equation}
  \label{eq:model1}
h_{\sigma\sigma'\tau}(\mathbf{k}) = d_{\tau,0}({\bf k}) \delta_{\sigma\sigma'}+ {\bf d}_\tau({\bf k}) \cdot \boldsymbol{\sigma}_{\sigma\sigma'}
\end{equation}
is an effective two-band Hamiltonian at node $\tau$ ($\tau\in\{1,2,3,4\}$) and $\boldsymbol{\sigma}$ is a vector of Pauli matrices, with $\sigma_z$ denoting the two bands at their touching point.
The sum over ${\bf k}$ in Eq.~(\ref{eq:H0}) is constrained by a cutoff that is assumed to be small compared to internodal distances.

For the $\tau=1$ node, we take
\begin{align}\label{eq:model2}
d_{1,0}(\mathbf{k}) &= \hbar v_F {\bf t}\cdot{\bf k} \nonumber \\
{\bf d}_1({\bf k}) & = \hbar v_F {\bf k},
\end{align}
where $v_F$ is the Fermi velocity and ${\bf t}$ is a dimensionless vector describing the magnitude and direction of the tilt of the Weyl cone.
Untilted Weyl cones are characterized by $t=0$, while a transition between type I and type II WSM\cite{Carbotte2016, Trescher2017, Tchoumakov2016} takes place at $t=1$.
Hereafter, we will focus on type I materials, where $t < 1$.
The use of an isotropic Fermi velocity can be justified for generic Weyl nodes by an appropriate rescaling of momenta.\footnote{At nonzero magnetic field, this involves a rescaling of the magnetic field as well. At any rate, the main results from this work are insensitive to the rescaling.}

In a real WSM, Eq.~(\ref{eq:model1}) holds only in a coordinate system made of principal axes, whose orientation with respect to the crystallographic axes can be rather complicated.\cite{Yu2016}
By assuming that the mirror plane is perpendicular to a principal axis ($k_z$ in our case), we are operating within the confines of a toy model.

The low-energy Hamiltonian for the other three Weyl nodes can be obtained by applying the mirror and time-reversal operations to $h_1({\bf k})$, according to the recipe
\begin{align}\label{eq:symtilt}
&\tau=1 \rightarrow 2   :  (v_F,t_x, t_y, t_y) \rightarrow (-v_F,-t_x, -t_y, t_z)\nonumber \\
&\tau=1 \rightarrow 3   :  (v_F,t_x, t_y, t_z) \rightarrow (v_F, -t_x, -t_y, -t_z)\nonumber \\
&\tau=1 \rightarrow 4   :  (v_F,t_x, t_y, t_z) \rightarrow (-v_F,t_x, t_y, -t_z),
\end{align}
where we have assumed that $\sigma_i$ transforms in the same way as angular momentum under time reversal and mirror operations.

This assumption requires some ellaboration.
In a real WSM, if the spin quantization axis is chosen along one of the crystallographic axes (say, the $c$ axis), then electronic eigenstates at a given Weyl node are linear combinations of spin-up and spin-down.
This is a consequence of spin-orbit interactions.
Accordingly, $\sigma_i=\sum_j a_{i j} s_j$, where $s_i$ are the true spin operators and $a_{i j}$ are real coefficients, for $i,j=x,y,z$.
Because all $s_i$ are odd under time reversal and all $a_{i j}$ are real, it follows that $\boldsymbol{\sigma}$ transforms like ${\bf s}$ under time reversal.
In contrast, $\boldsymbol{\sigma}$ and ${\bf s}$ do not in general transform in the same way under the mirror operation, because (i) $\sigma_i$ is a linear combination of different $s_j$, and (ii) the components of ${\bf s}$ parallel to the mirror plane change sign under the mirror operation, whereas the component perpendicular to the mirror remains invariant.
Thus, it is mainly for the sake of analytical simplicity that we assume the transformation rules of Eq.~(\ref{eq:symtilt}).
Nonetheless, the main predictions based on our toy model will be general enough to trascend its simplicity.

\begin{figure}[t]
\includegraphics[width=0.45 \textwidth]{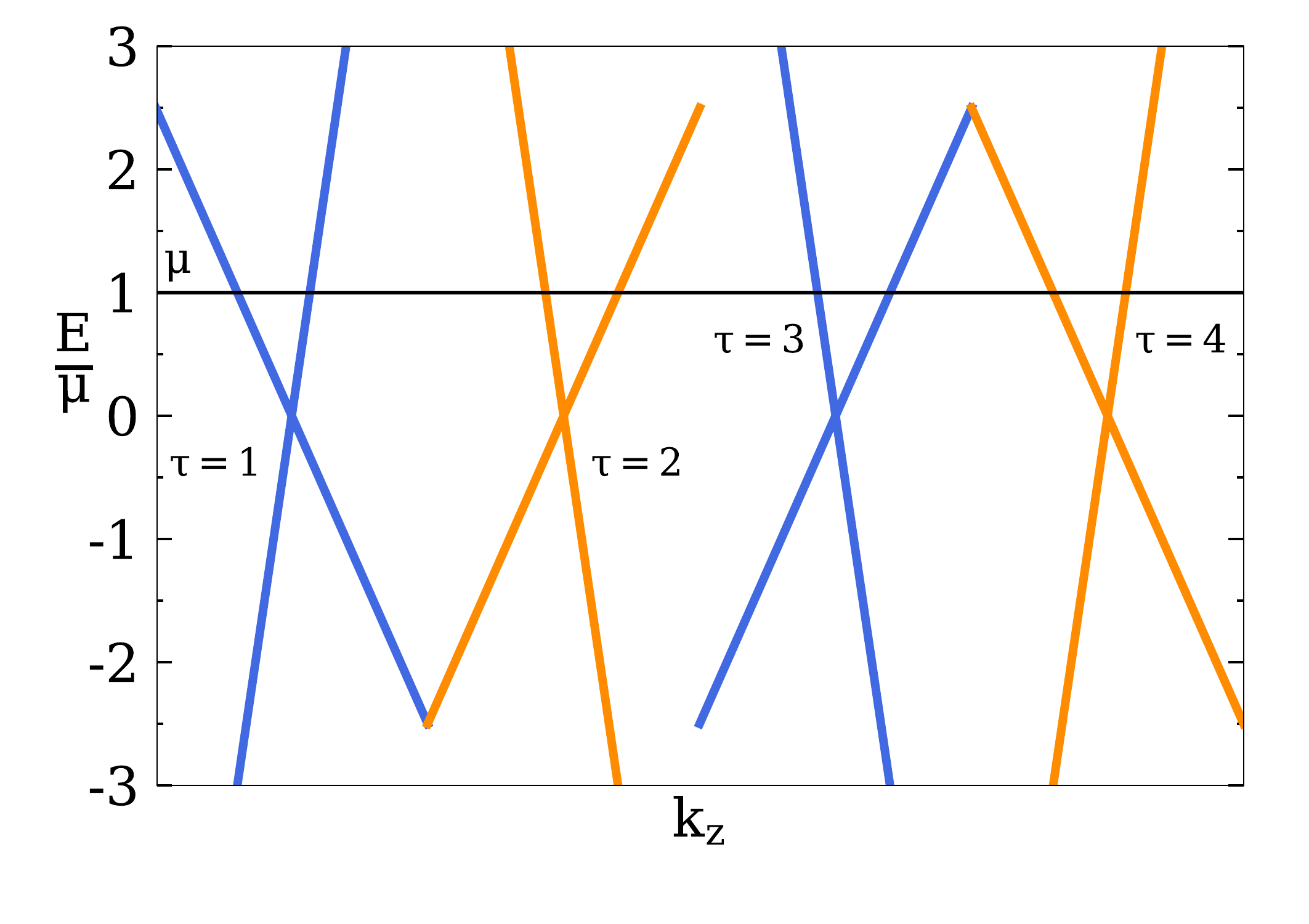}
\caption{
  Energy dispersion of the four-node model for a finite tilt in the absence of magnetic field and Coulomb interactions, for $k_x=k_y=0$.
  For clarity, the positions of different nodes have been shifted in $k_z$. The two different colors represent the two chiralities. }
\label{fig:disp_zero_field}
\end{figure}

The energy spectra for all nodes are displayed in Fig.~\ref{fig:disp_zero_field}.
Nodes $1$ and $3$ are partners under time-reversal, as are nodes $2$ and $4$.
The two pairs are related by the mirror plane.

At vanishing tilt, each node displays an antiunitary symmetry: $h_\tau({\bf k}) = \Theta^{-1} h_\tau(-{\bf k}) \Theta$, where $\Theta = i \sigma_y K$ and $K$ is the complex conjugation.
This symmetry, which is broken when ${\bf t}\neq 0$,  is not the true time-reversal symmetry.
The latter connects different Weyl nodes and is unbroken by a tilt.
Also, $\Theta$ is not equivalent to the particle-hole symmetry.
A tilt of the Weyl cone happens to break both $\Theta$ and particle-hole symmetry, but nonlinear terms of the type used in Ref. [\onlinecite{Bertrand2017}] break $\Theta$ while preserving particle-hole symmetry.

\begin{figure}[h!]
  \includegraphics[width=0.45 \textwidth]{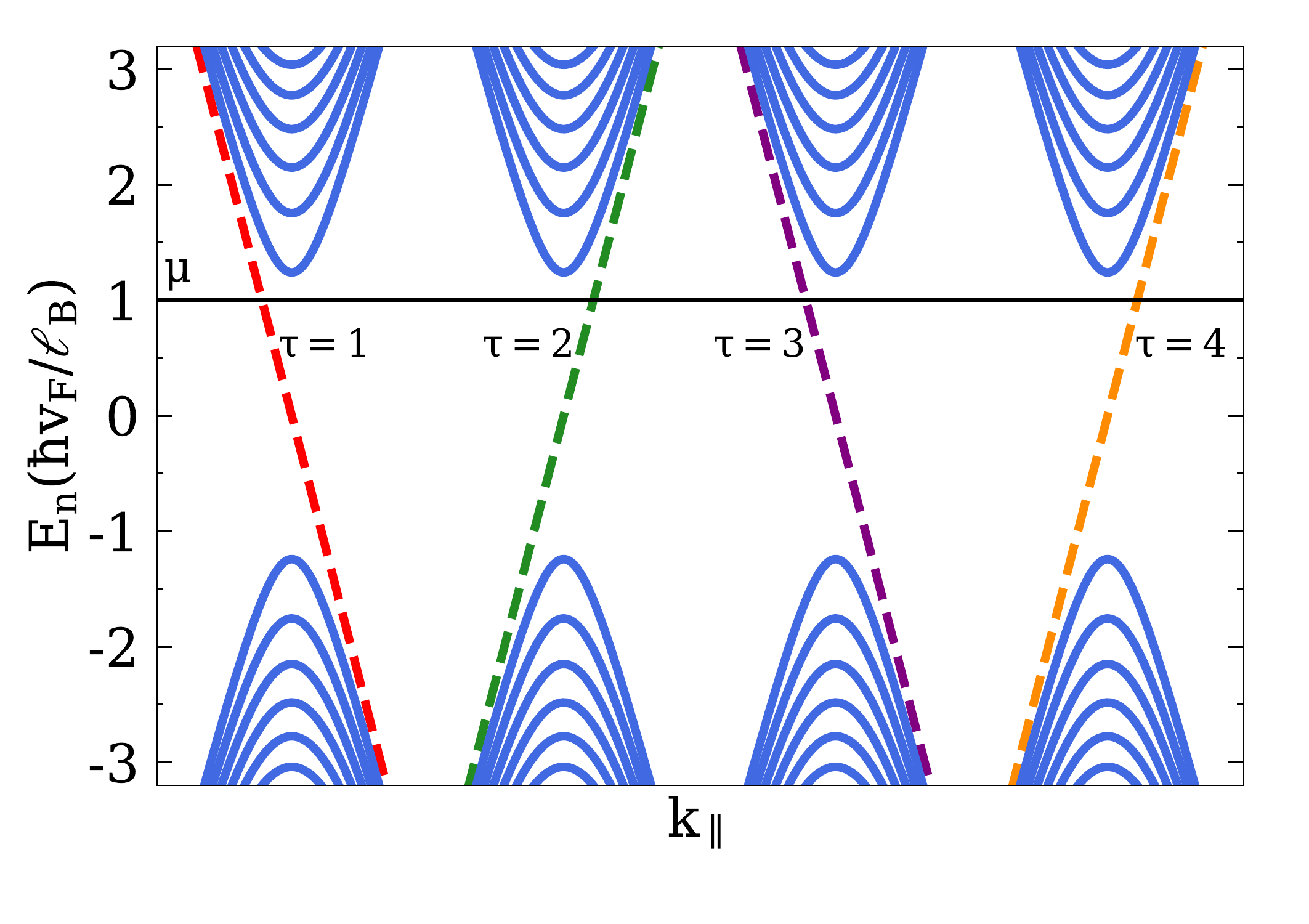}
  \includegraphics[width=0.45 \textwidth]{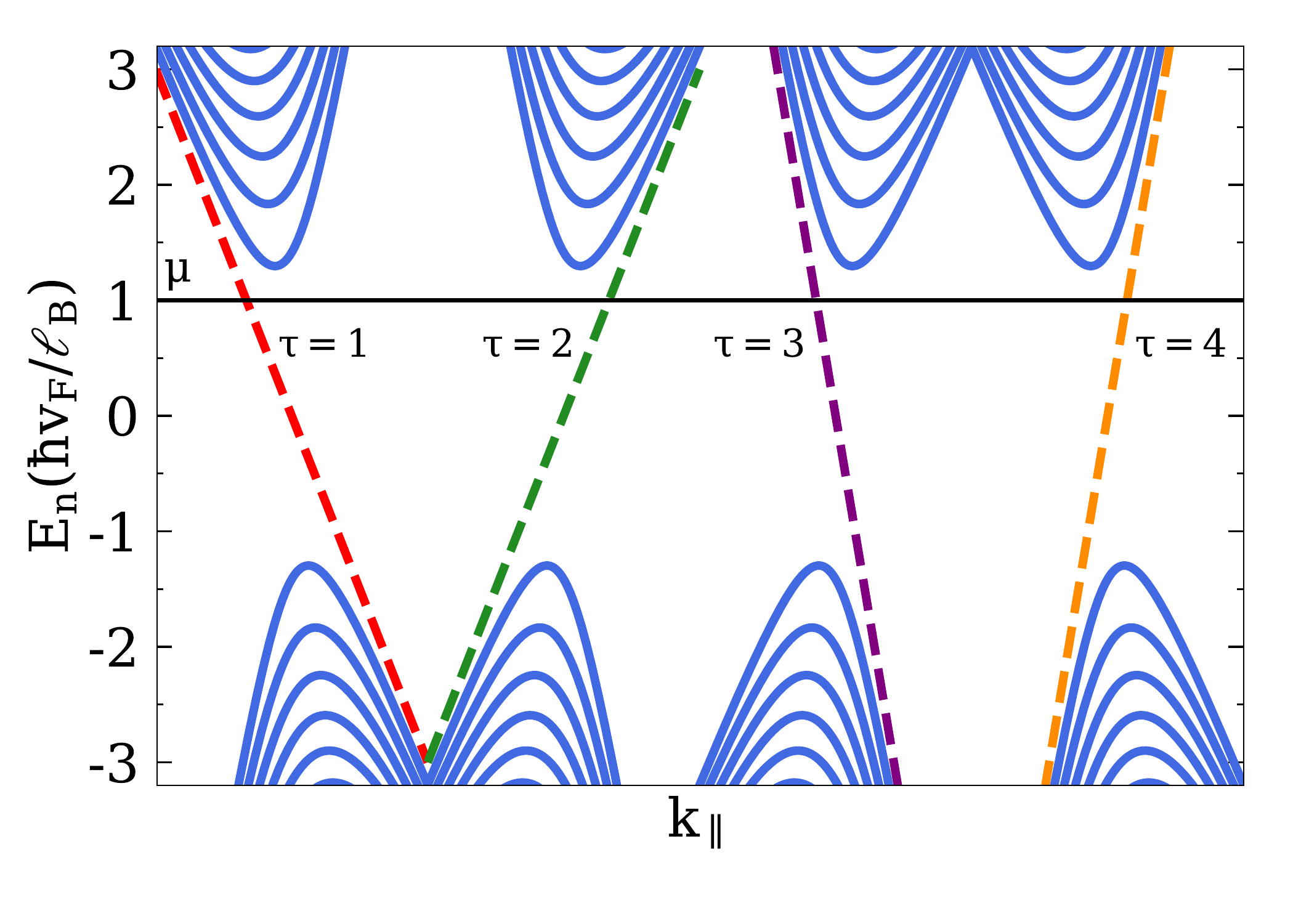}
  \caption{
    Energy dispersion of the four-node WSM model in a magnetic field, without Coulomb interactions.
    The tilt vector is chosen along $z$.
    In the top panel, the magnetic field is oriented perpendicularly to the tilt vectors of all four nodes.
    Consequently, the optical gaps are identical for all the nodes.
    In the bottom panel, the magnetic field is perpendicular to the mirror plane and has therefore a nonzero projection along the tilt vectors.
    In this case, the optical gaps for interband transitions involving the chiral Landau level are different for time-reversal node partners, but equal for mirror partners.}
  \label{fig:disp_nonzero_field}
\end{figure}

Next, we add a uniform and static magnetic field ${\bf B}=B\hat{\bf B}$ through the minimal substitution ${\bf k}\to {\bf k} - e {\bf A}/\hbar $, where ${\bf A}$ is the vector potential.
In the absence of electron-electron interaction, the electronic dispersion on each Weyl node can be found analytically.\cite{Tchoumakov2016}
We will label the energy levels with an integer $n$, with $n>0$ ($n<0$) for the conduction (valence) nonchiral Landau levels, and $n=0$ for the chiral Landau level.
With this notation, the energy levels for the $\tau=1$ node are given by
\begin{align}
  E_{n\neq 0, k_\parallel}^{(1)}  &=  \hbar v_F t_\parallel k_\parallel  + {\rm sgn}(n) \frac{\hbar}{\gamma}\sqrt{v_F^2 k_\parallel^2 + \frac{2 |n| v_F^2}{\gamma \ell_B^2}}\nonumber\\
  E_{n=0, k_\parallel}^{(1)}  &= \hbar v_F t_\parallel k_\parallel - \frac{\hbar v_F}{\gamma} k_\parallel,
\label{eq:tchouma}
\end{align}
where $k_\parallel = {\bf k}\cdot\hat{\bf B}$ is the wave vector in the direction parallel to the magnetic field, $t_\parallel = {\bf t}\cdot\hat{\bf B}$ is the projection of the tilt along the magnetic field, $l_B=\sqrt{\hbar/(e B)}$ is the magnetic length,
and $\gamma = (1-|{\bf t}\times\hat{\bf B}|^2)^{-1/2}$.
It is worth emphasizing that Eq.~(\ref{eq:tchouma}) is valid for an arbitrary direction of ${\bf B}$.
 The spectrum for the remainder of the Weyl nodes can be rapidly obtained using Eq.~(\ref{eq:symtilt}).
Some examples of the electronic spectra are displayed in Fig.~\ref{fig:disp_nonzero_field}.

Throughout this work, we omit Zeeman terms of the type $\boldsymbol{\sigma}\cdot{\bf B}$ in the model Hamiltonian.
Such terms have been included in Ref.~[\onlinecite{Jiang2018}], along with an {\em ad hoc} value for the $g$ factor, in order to fit the experimental data.
In our toy model, Zeeman terms have two unimportant consequences.
First, there is a displacement in the positions of the Weyl nodes, which can be absorbed by redefining ${\bf k}$. 
Second, there is an overall energy shift of ${\bf t}\cdot{\bf B}$ for all Weyl nodes, which can be absorbed in the chemical potential $\mu$.

Thus far we have considered free fermions.
Coulomb interactions add a term
\begin{equation}
\label{eq:elel}
{\cal H}_{\rm el-el}\simeq\frac{1}{2\mathcal{V}}\sum_{\mathbf{q}} V({\bf q})\rho\left(\mathbf{q}\right)\rho\left(-\mathbf{q}\right)
\end{equation}
to the low-energy effective Hamiltonian, where ${\cal V}$ is the sample volume,  $V({\bf q}) = e^2/(\varepsilon_\infty {\bf q}^2)$ is the Coulomb potential and $\rho\left(\mathbf{q}\right)\equiv\sum_{{\bf k} \sigma\tau} c_{\mathbf{k}\sigma\tau}^{\dagger}c_{\mathbf{k}+\mathbf{q}\sigma\tau}$ is the low-energy electronic density operator.
The parameter $\varepsilon_\infty$ is the contribution from high-energy electronic states (not included in the minimal model) to the dielectric function, in units of the vacuum permittivity.
In writing Eq.~(\ref{eq:elel}), we have kept only the long-wavelength part of the Coulomb interaction, thereby neglecting internode Coulomb scattering.

\subsection{Optical conductivity}

The optical absorption is determined by the real part of the optical conductivity $\sigma_{\alpha \beta}$,
\begin{equation}\label{eq:optcond}
{\rm Re} \left[\sigma_{\alpha \beta}(\omega)\right]= - \lim_{q\rightarrow 0} \frac{{\rm Im}\left[\chi^R_{J_\alpha J_\beta}(\mathbf{q},\omega)\right]}{\omega},
\end{equation}
where $\alpha, \beta\in\{x,y,z\}$, ${\bf q}$ and $\omega$ are the wave vector and frequency of the incoming photons,  and $\chi^R_{J_\alpha J_\beta}(\mathbf{q},\omega)$ is the retarded current-current response function.
Although the magnitude of ${\bf q}$ is approximated as zero, its direction matters because the electromagnetic vector potential for the incident light, which couples to the velocity operator of Weyl fermions, is perpendicular to $\hat{\bf q}$.

We evaluate $\chi$ using the low-energy Hamiltonian presented above and treating the Coulomb interactions within the generalized random-phase approximation.
We refer the reader to Appendix \ref{App_GRPA} for details about the formalism and calculations.

The neglect of internode Coulomb scattering, alluded to above,  enables us to write
\begin{equation}
 \sigma_{\alpha \beta}(\omega)=\sum_\tau \sigma_{\alpha \beta}(\omega,\tau),
\end{equation}
where $\sigma_{\alpha \beta}(\omega,\tau)$ is the conductivity corresponding to an isolated Weyl node $\tau$.
In this work, we will be interested in the optical response to left- and right-handed circularly polarized photons (LCP and RCP, respectively).
These responses can be immediately obtained from $\sigma_{\alpha\beta}$. For instance, if ${\bf q}||\hat{\bf z}$,
\begin{align}\label{eq:lcprcp}
\sigma_{\rm LCP}(\omega) &= \sigma_{xx}(\omega) - i \sigma_{xy}(\omega)\nonumber \\ 
\sigma_{\rm RCP}(\omega) &= \sigma_{xx}(\omega) + i \sigma_{xy}(\omega).
\end{align}
The two conductivities are related via the condition $\sigma_{\text{LCP},{\bf B}}(\omega,\tau) = \sigma_{\text{RCP},-{\bf B}}(\omega,\bar{\tau})$, where the node $\bar{\tau}$ is the time-reversed partner of node $\tau$.
This in turn implies $\sigma_{\text{LCP},{\bf B}}(\omega) = \sigma_{\rm RCP,-{\bf B}}(\omega)$ for the total optical conductivity.

\section{Results}\label{sec:res}

This section presents the results for the optical absorption, calculated from the toy model of Sec.~\ref{sec:mod}.
We concentrate on the zero-temperature interband absorption, thereby omitting the Drude peak. The main statements of this section hold irrespective of the presence of Coulomb interactions, the role of which will be made explicit at the end of the section.
In Sec.~\ref{sec:Disc}, we will examine the applicability of our results to real WSM.

\subsection{Optical absorption and valley polarization}
\label{sec:res1}

For pedagogical purposes, we begin with a discussion of the optical absorption in the absence of a magnetic field.
This discussion extends that of Ref.~[\onlinecite{Bertrand2017}], where we focused on a particle-hole symmetric energy dispersion with linear and nonlinear terms in the electronic momentum.
As shown in Appendix \ref{App_nonlin}, the effect of a nonlinear energy dispersion in the optical absorption is qualitatively similar to that of a tilt.
Here we consider a tilted and linear Weyl dispersion, as it is analytically simpler.

In the case of a vanishing tilt, the optical absorption is identical for all four nodes regardless of the direction of ${\bf q}$.
This can be understood from a symmetry point of view.
First, we recognize that circularly polarized light reverses polarization under complex conjugation, and transforms as a pseudovector under operations belonging to the crystalline point group. 
Then, the effective antiunitary symmetry mentioned in Sec.~\ref{subsec:model} implies that $\sigma_{\rm LCP}(\omega,\tau)=\sigma_{\rm RCP}(\omega,\tau)$.
Second, the true time-reversal symmetry at zero magnetic field implies that $\sigma_{\rm LCP}(\omega,\tau) = \sigma_{\rm RCP}(\omega,\overline{\tau})$.
Putting these two conditions together, we have $\sigma_j(\omega,\tau) = \sigma_j(\omega,\overline{\tau})$, $j$ being either LCP or RCP.
Third, we have $\sigma_j(\omega,1) = \sigma_j(\omega,2)$ and $\sigma_j(\omega,3) = \sigma_j(\omega,4)$.
These relations are ensured by the mirror symmetry (if the light propagates perpendicularly to the mirror plane) or by the product of mirror and time-reversal operations (if the light propagates parallel to the mirror).

\begin{figure}[t]
  \includegraphics[width=0.45 \textwidth]{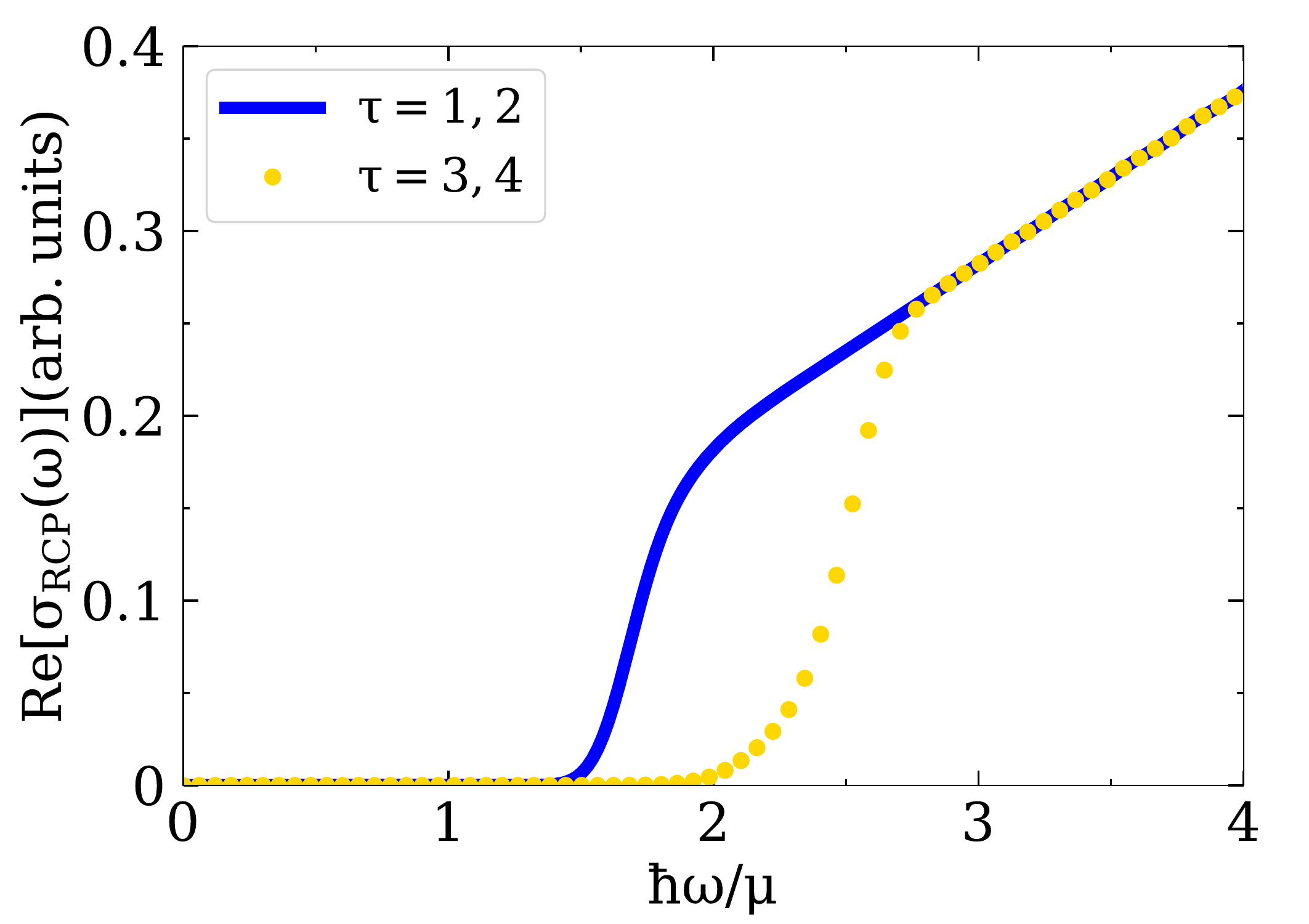}
\caption{
  Node-resolved interband optical absorption for the minimal four-node model, in the absence of a magnetic field and without Coulomb interactions.
  We assume a tilt vector ${\bf t}=(0.2,0,0.4)$ (i.e., neither perpendicular nor parallel to the mirror).
  The momentum ${\bf q}$ of the circularly polarized light is oriented perpendicular to the mirror plane.
  Time-reversed partner nodes ($1$ and $3$, or $2$ and $4$) have different absorption intensities on a finite frequency range.
  However, the optical gaps are the same (the absorption starts at the same frequency) for all nodes.}
  \label{fig:abs_nofield}
\end{figure}

\begin{figure}[t]
\includegraphics[width=0.45 \textwidth]{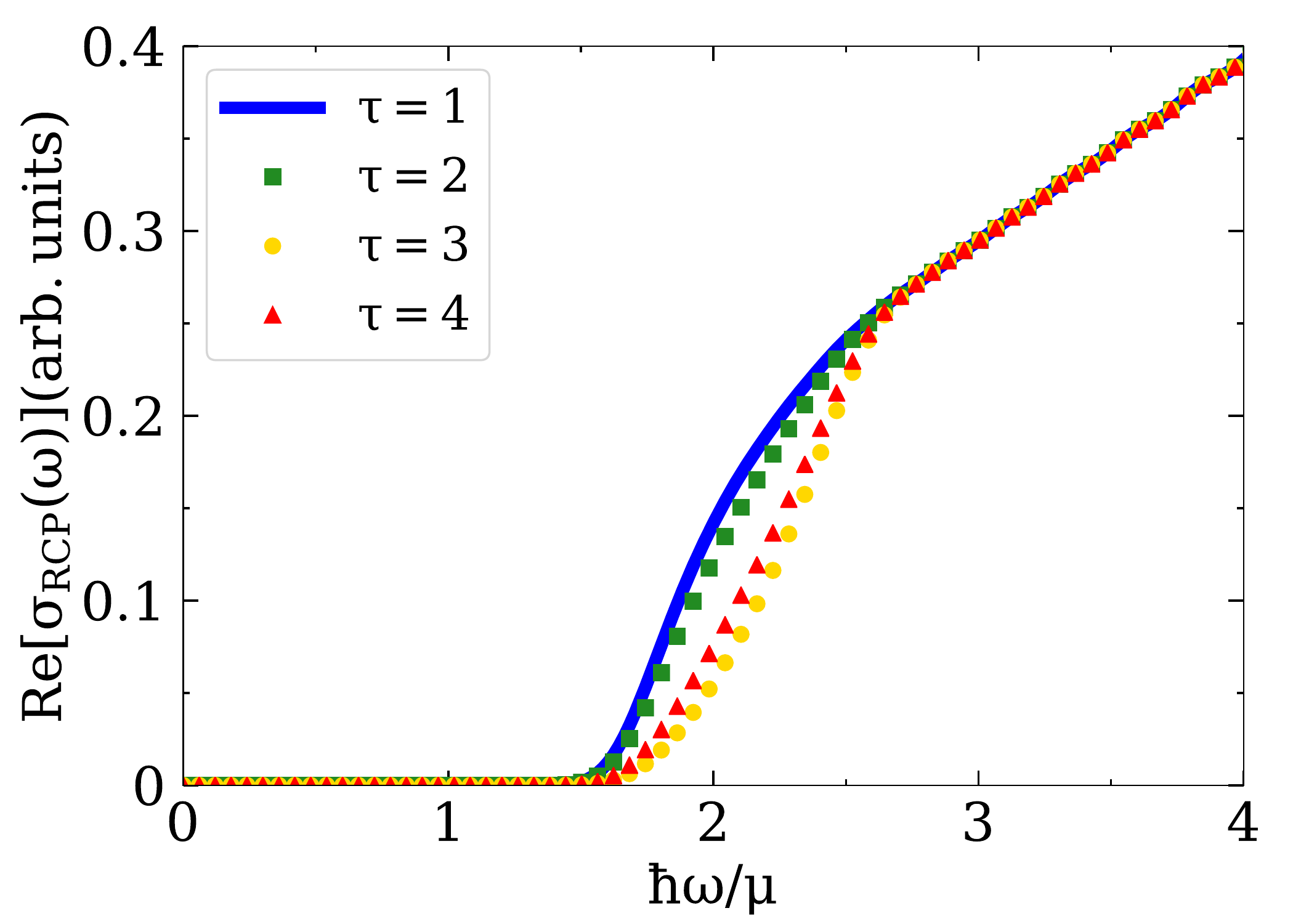}
\caption{
  Node-resolved interband optical absorption for the minimal four-node model, in the absence of a magnetic field and without Coulomb interactions.
  We assume a tilt vector ${\bf t}=(0.2,0,0.4)$ (i.e., neither perpendicular nor parallel to the mirror).
  The momentum ${\bf q}$ of the circularly polarized light makes an angle of $\pi/6$ with the $z$ axis.
  In this configuration, all four nodes display a different optical absorption spectrum.
  However, the optical gaps are the same for all nodes.}
  \label{fig:abs_nofield2}
\end{figure}

A tilt of the Weyl cones breaks the antiunitary symmetry $\Theta$ (much like a nonlinearity in the energy spectrum does) and thus enables a difference in the optical absorption at different nodes.
Assuming the generic situation in which the direction of the tilt vector is neither parallel nor perpendicular to the mirror plane, there are three main possibilities.

First, as shown in Fig.~\ref{fig:abs_nofield}, if ${\bf q}$ is perpendicular to the mirror plane, the pair of nodes $(1,2)$ absorbs light with different intensity than the pair of nodes $(3,4)$.
This constitutes a pairwise valley polarization.
Reversing the direction of ${\bf q}$ (or equivalently switching RCP$\leftrightarrow$LCP) exchanges the absorption intensity of the two pairs of nodes. 
The reason why nodes related by mirror symmetry ($1$ and $2$,  or $3$ and $4$) absorb light equally is that circularly polarized light preserves the mirror symmetry when directed perpendicularly to the mirror plane.

Second, if ${\bf q}$ is parallel to the mirror plane, nodes $(1,4)$ absorb light differently from nodes $(2,3)$.
In this configuration, the circularly polarized light breaks the mirror symmetry and thus Weyl nodes related by a mirror reflection do not display an equal absorption intensity.
Nevertheless, the configuration preserves the product of time reversal and mirror symmetries, which is the reason why nodes $1$ and $4$ (or $2$ and $3$) absorb light equally.

Third, for the most general case in which ${\bf q}$ is neither parallel nor perpendicular to the mirror plane, all four nodes present a different optical absorption (see Fig. \ref{fig:abs_nofield2}).

Though non generic, it may happen that ${\bf q}$ is simultaneously perpendicular to the tilt vectors of all Weyl nodes.
This circumstance applies to the so-called W1 Weyl nodes in TaAs and related materials, as will be discussed in Sec. \ref{sec:Disc}.
For that case, we find that all nodes absorb the light equally, as though there were no tilt at all.

\begin{table}[tb]
  \caption{Optical (interband) valley polarization at zero magnetic field in a time-reversal symmetric Weyl semimetal with four tilted Weyl cones and a mirror plane. The tilt vector for the $\tau=1$ Weyl node is ${\bf t}$, and the circularly polarized incident light has a momentum ${\bf q}$. The notation $(n,m)$ means that nodes $n$ and $m$ absorb light with equal intensity. We consider the generic case in which ${\bf t}$ has nonzero components parallel and perpendicular to the mirror plane.}
  \label{tab:vp1}
\begin{ruledtabular}
\begin{tabular}{cc}
Configuration & Valley polarization  \\ \hline
${\bf q} \perp$ mirror & Partial, $(1,2) \neq (3,4)$ \\ \hline
${\bf q} ||$mirror  & Partial, $(1,4) \neq (2,3)$ \\ \hline
${\bf q} \not\perp,\not\parallel$ mirror & Partial, $1\neq 2\neq 3 \neq 4$ 
\end{tabular}
\end{ruledtabular}
\end{table}

Table ~\ref{tab:vp1} summarizes the preceding statements: in a WSM with tilted cones, a valley polarization is obtained at zero magnetic field under irradiation by circularly polarized light (we have checked that linearly polarized light produces no valley polarization at zero field).
The origin of the valley polarization lies on the interband matrix elements of the current operator.\cite{Bertrand2017,Yu2016}
In all cases, the optical gap (i.e., the minimal frequency for the onset of interband light absorption) is the same for all four nodes.
This fact prevents the valley polarization from attaining $100$\% (hence the label ``partial'' in Table~\ref{tab:vp1}) and makes the experimental detection challenging.
Moreover, we note that the valley polarization is always nonchiral, i.e.  $\sigma_j(\omega,1)+\sigma_j(\omega,3) = \sigma_j(\omega,2)+\sigma_j(\omega,4)$ for $j=$LCP, RCP.

\begin{table}[tb]
    \caption{Optical gaps for inter Landau level transitions involving the chiral Landau level, in a time-reversal symmetric Weyl semimetal with four tilted Weyl cones and a mirror plane. The notation $(n,m)$ means that nodes $n$ and $m$ have the same optical gaps. In this table, the tilt is assumed to have nonzero components parallel and perpendicular to the mirror plane.}
  \label{tab:opgap}
\begin{ruledtabular}
\begin{tabular}{ccc}
Configuration & Preserved symmetry& Optical gaps  \\ \hline
${\bf B} \perp$ mirror &  mirror & $(1,2) \neq (3,4)$  \\ \hline
${\bf B} ||$ mirror & (mirror) $\times$ (time-reversal) &$(1,4) \neq (2,3)$ \\ \hline
${\bf B} \not\parallel,\not\perp$ mirror & none & $1\neq 2\neq 3 \neq 4$ 
\end{tabular}
\end{ruledtabular}
\end{table}

\begin{table}[tb]
  \caption{Optical (interband) valley polarization at magnetic field ${\bf B}$ in a time-reversal symmetric Weyl semimetal with four tilted Weyl cones and a mirror plane.
We consider the generic case in which tilt vectors are neither perpendicular nor parallel to the mirror plane.
    The circularly polarized incident light has a momentum ${\bf q}$. The notation $(n,m)$ means that nodes $n$ and $m$ absorb light with equal intensity. The label ``full'' refers to a complete valley polarization for a finite frequency range of the incident light.
The notation $[n,m]$ means that nodes $n$ and $m$ have the same optical gaps but absorb light with different intensities, thereby producing a partial valley polarization.}
  \label{tab:vp2}
\begin{ruledtabular}
\begin{tabular}{cc}
Configuration &  Valley polarization \\ \hline
${\bf B} \perp$ mirror, ${\bf q} || {\bf B}$ & Full, $(1,2) \neq (3,4)$   \\ \hline
${\bf B} \perp$ mirror, ${\bf q} \not\parallel {\bf B}$  & Full, $[1,2] \neq [3, 4]$   \\ \hline
${\bf B} ||$ mirror, ${\bf q} || {\bf B}$ & Full, $(1,4) \neq (2,3)$   \\ \hline
${\bf B} ||$ mirror, ${\bf q} \not\parallel {\bf B}$  & Full, $[1,4] \neq [2, 3]$ \\ \hline  
${\bf B} \not\parallel, \not\perp$ mirror, all ${\bf q}$ & Full, $1\neq 2\neq 3\neq 4$
\end{tabular}
\end{ruledtabular}
\end{table}

The behavior of the optical absorption becomes richer and more interesting in the presence of a magnetic field.
For one thing, there are more configurations to consider depending on the relative orientations between the mirror plane, ${\bf t}$, ${\bf B}$ and ${\bf q}$. \footnote{The momentum separation between Weyl nodes does not enter in the consideration of our low-energy model. The predictions from this model are expected to be approximately valid when the inverse of the magnetic length is small compared to the wave vector separation between the Weyl nodes.}
Another novelty brought about by the magnetic field is that the optical gaps can become different for Weyl nodes that were symmetry-equivalent in the absence of the field (see Table \ref{tab:opgap} and Fig.~\ref{fig:disp_nonzero_field}).
This feature, which follows from Eq.~(\ref{eq:tchouma}), opens the way for a complete ($100\%$) valley polarization at strong magnetic fields.
An exception takes place in the non generic case where the magnetic field is simultaneously perpendicular to the tilt vectors of all the Weyl nodes.
Then, by virtue of Eq.~(\ref{eq:tchouma}), the optical gaps become equal for all Weyl nodes.

\begin{figure}[t]
\includegraphics[width=0.45 \textwidth]{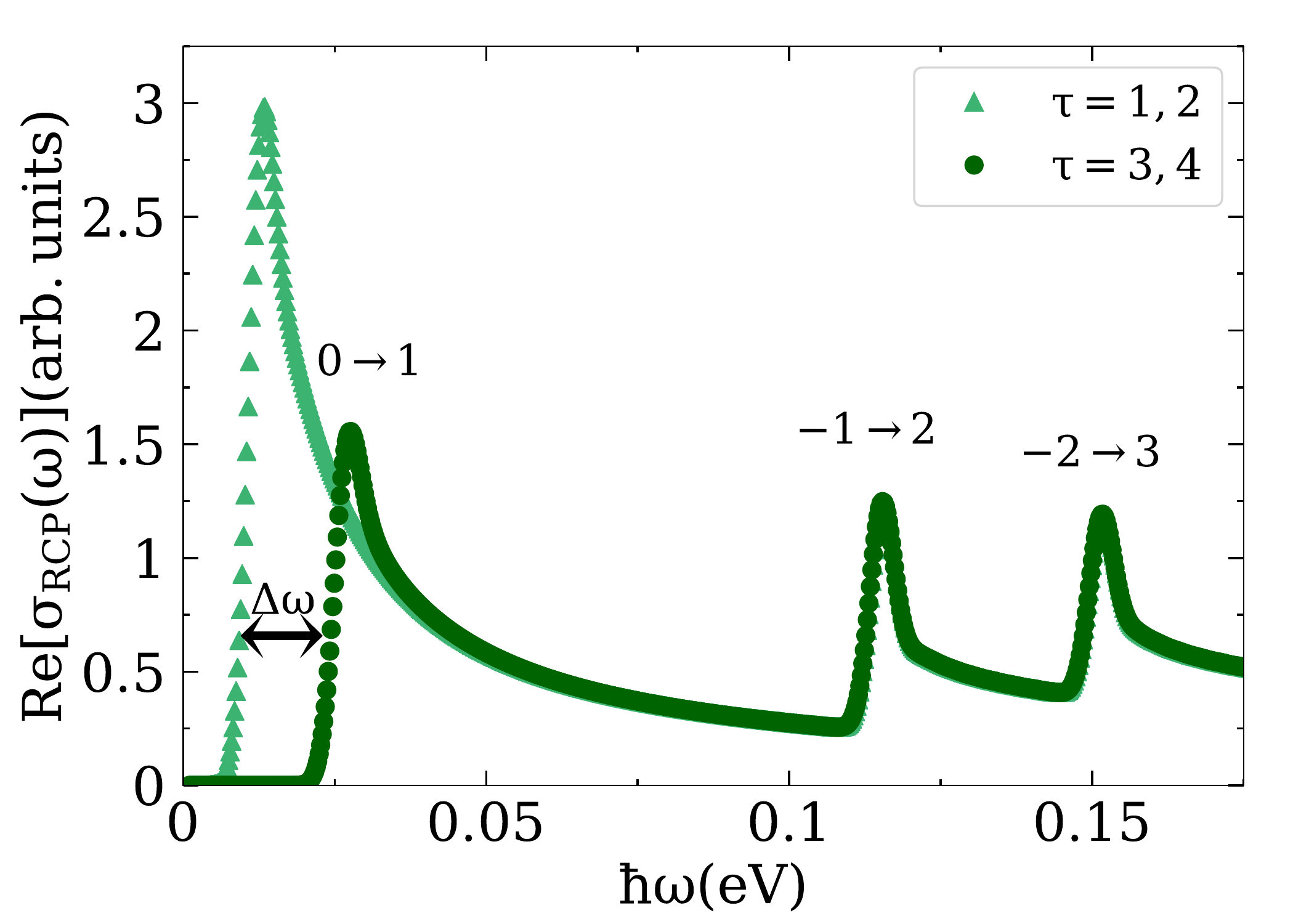}
\caption{
  Node-resolved interband optical absorption for a four-node Weyl semimetal with Coulomb interactions, in a strong magnetic field. 
  The right-circularly-polarized incident light propagates along $z$, and the magnetic field is also directed along $z$.
  The tilt vector for the first Weyl node is chosen as ${\bf t}=(0,0,0.5)$.
  The parameter values are as follows: $B=20\text{T}$ for the magnetic field, $\epsilon_\infty = 25$ for the high-frequency dielectric constant, and $\mu = \hbar v_F / \ell_B$ for the chemical potential.
  The broadening of the Lorentzian in the expression for the optical absorption is taken to be $\eta = 2\text{meV}$. 
  In this configuration, the allowed interband transitions are $0 \to 1$, $-1 \to 2$, $-2 \to 3$; these transitions are indicated in the figure.
  Time-reversed partner nodes ($1$ and $3$, or $2$ and $4$) have different optical gaps, which leads to a frequency interval $\Delta\omega$ where valley polarization becomes complete.}
  \label{fig:abs_field}
\end{figure}

\begin{figure}[h!]
\includegraphics[width=0.45 \textwidth]{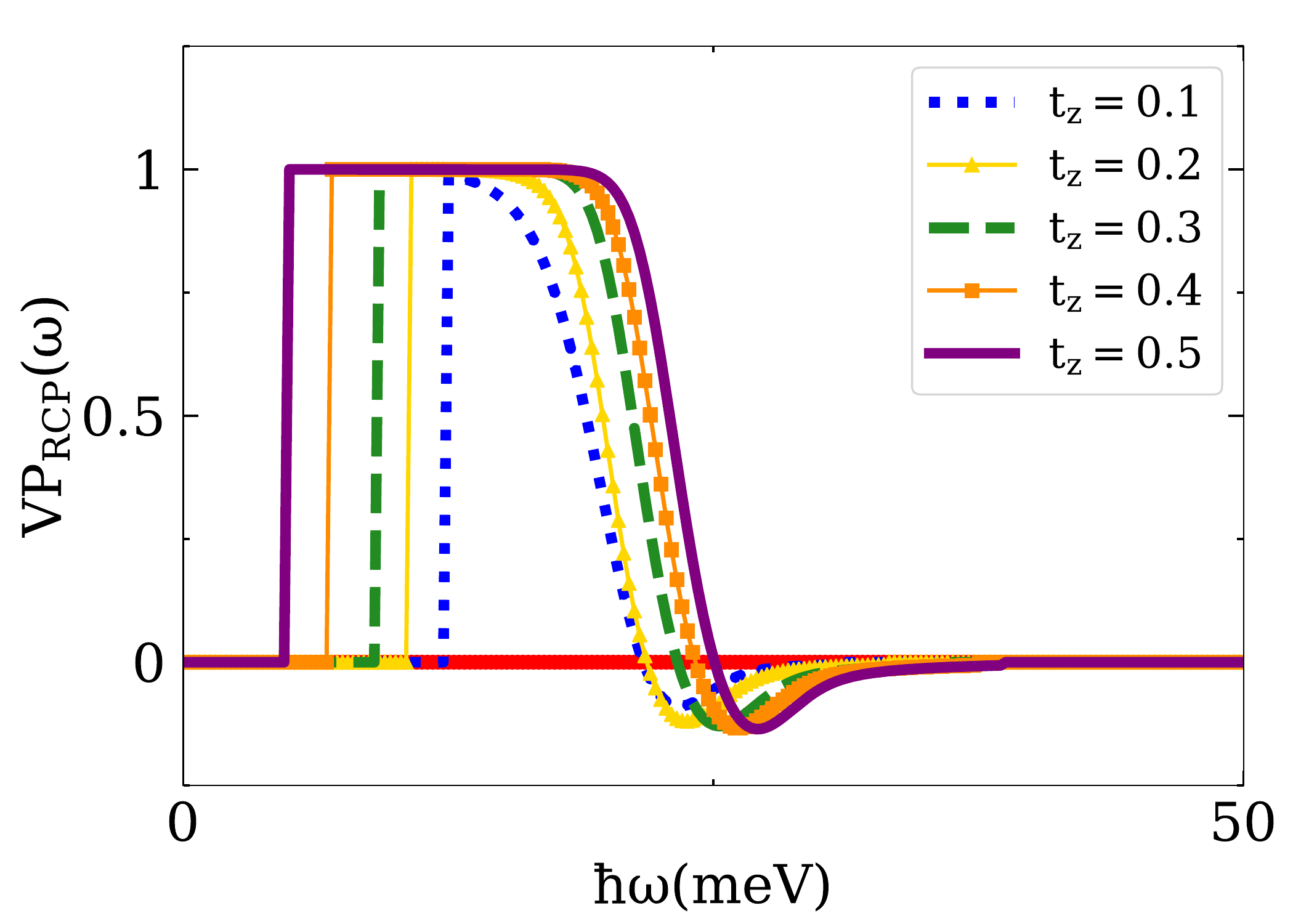}
\includegraphics[width=0.45 \textwidth]{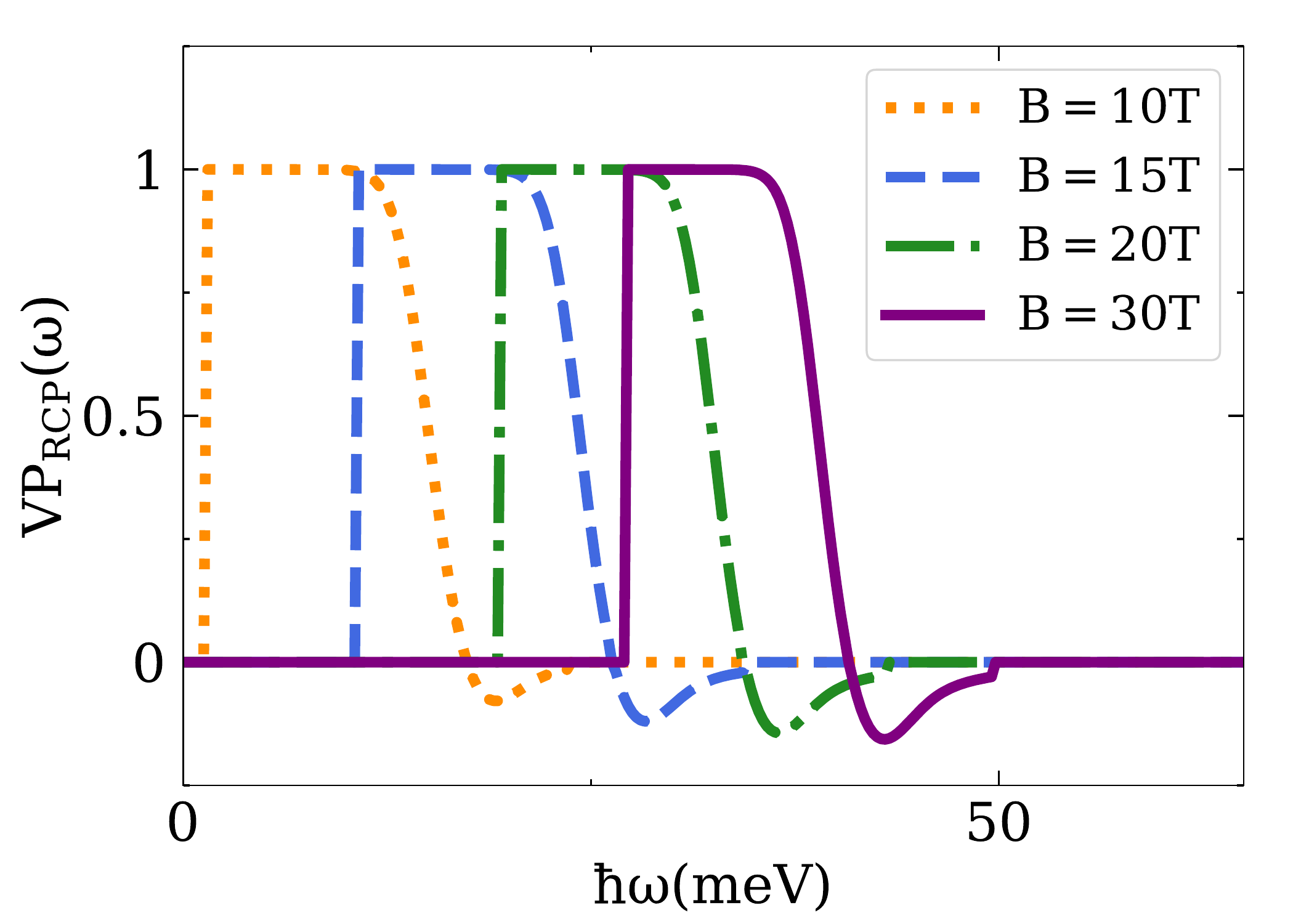}
\caption{
  Frequency-dependent valley polarization (Eq.~(\ref{eq:valpol})) in a four-node Weyl semimetal with Coulomb interactions.
  Unless otherwise stated, the $({\bf q},{\bf B}, {\bf t})$ configuration and parameter values are the same as in Fig.~\ref{fig:abs_field}.
  (Top panel) The frequency interval for the complete valley polarization increases with the tilt ($B=20 {\rm T}$ in this panel).
  (Bottom panel) The frequency interval for the complete valley polarization decreases as the magnetic field drives the system deeper into the quantum limit .}
  \label{fig:valpol_tilt}
\end{figure}

The results for the node-resolved optical absorption are summarized in Table~\ref{tab:vp2}.
For instance, if ${\bf B}$ and ${\bf q}$ are parallel to each other but perpendicular to the mirror plane, the optical gap for interband transitions involving the chiral LL becomes different for Weyl nodes related by time reversal.
Consequently, in the quantum limit (where the Fermi level intersects only the chiral LLs), there exists a finite frequency range for which circularly polarized photons are absorbed at nodes $\tau=1$ and $\tau=2$, but not at nodes $\tau=3$ and $\tau=4$ (see Fig.~\ref{fig:abs_field}).
In other words, a 100\% valley polarization is attained in such frequency range.
To illustrate this, Fig. \ref{fig:valpol_tilt} shows the frequency-dependent valley polarization under a RCP incident light that propagates in the direction parallel to ${\bf B}$.
In this configuration, the valley polarization is 
\begin{align}\label{eq:valpol}
  \text{VP}_j(\omega) &= \frac{{\rm Re} \left[\sigma_j(\omega,\tau=1)\right]-{\rm Re} \left[\sigma_j(\omega, \tau=3)\right]}{{\rm Re} \left[\sigma_j(\omega,\tau=1)\right]+{\rm Re} \left[\sigma_j(\omega,\tau=3) \right]},
\end{align}
where $j=$RCP.
This valley polarization, originating from the $0\to 1$ inter LL transition, reaches 100\% (VP $=1$) for a significant frequency interval $\Delta\omega$ of the incident light ($\Delta\omega\simeq 15 \text{meV}$ when $|t_\parallel| = 0.5$, $B= 20 {\rm T}$ and $\mu \simeq \hbar v_F/l_B$).
For fixed ${\bf q}$, changing ${\bf B}\to-{\bf B}$ yields ${\rm VP=1}$ in a different frequency window.
For fixed ${\bf B}$, changing RCP into LCP (or ${\bf q}$ into $-{\bf q}$) yields ${\rm VP}=-1$, originating from the $-1\to 0$ transition, in a different frequency window.
We note incidentally that, in a magnetic field, a nonzero valley polarization also takes place under irradiation by linearly polarized light. 

The valley polarization can be further controlled by tuning ${\bf q}$ and ${\bf B}$.
For example, when ${\bf q}$ and ${\bf B}$ are perpendicular to the mirror plane, a simultaneous reversal of the field and of the light propagation direction (or its circular polarization) results in a change of the pair of nodes absorbing the light from $(1,2)$ to $(3,4)$, at a fixed frequency.
Similarly, by aligning ${\bf B}$ and ${\bf q}$ with each other and with the mirror plane, the pair of nodes that absorb the light switches from $(1,2)$ to $(1,4)$ (or from $(3,4)$ to $(2,3)$), at a fixed frequency.
The reason why nodes $1$ and $4$ (or $2$ and $3$) show an identical optical absorption intensity when ${\bf B}$ and ${\bf q}$ are aligned with each other and with the mirror plane is that the product of time-reversal and mirror symmetry is preserved in this configuration (even though each symmetry is individually broken).

Three other configurations are also worth mentioning.
First, when ${\bf q}$ and ${\bf B}$ are not collinear, all symmetries of the model Hamiltonian are broken, the four nodes display an unequal absorption intensity and the valley polarization between the pair of nodes having the same (different) optical gaps is partial (full). 

\begin{figure}[t]
\includegraphics[width=0.47\textwidth]{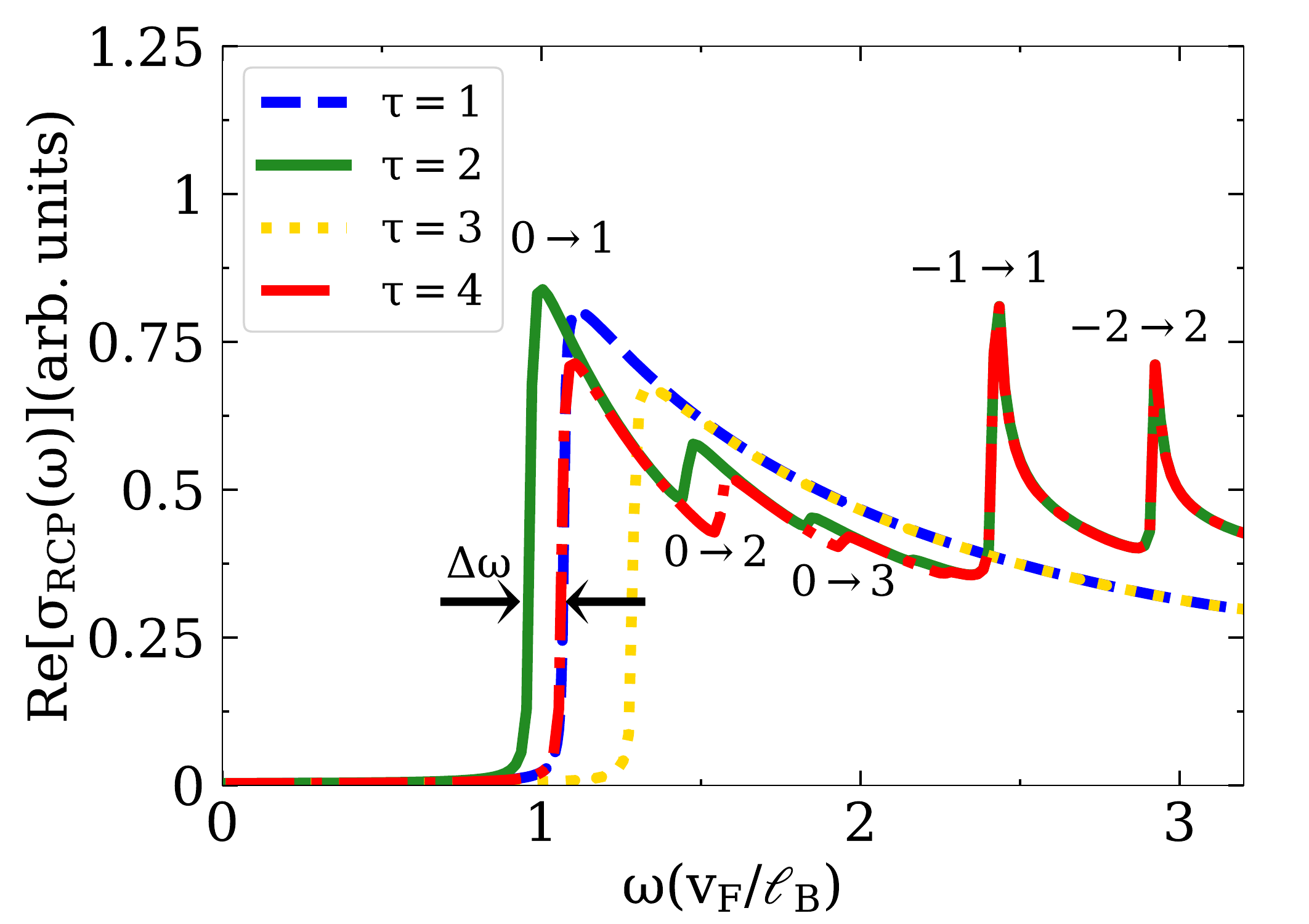}
\caption{
  Node-resolved interband optical absorption for the four-node Weyl semimetal.
  The magnetic field is directed parallel to the tilt vector of the $\tau=1$ Weyl ($\mathbf{t}=(0.25,0,0.43)$), and the RCP light propagates parallel to the magnetic field.
We take $\mu = 0.2 \, \hbar v_F /\ell_B$ and $\epsilon_\infty = \infty$ (non-interacting electrons).
In this configuration, there is a finite range of frequency $\Delta \omega \simeq 0.13 \hbar v_F/ \ell_B$, where only the $\tau=2$ node absorbs photons.
We have $\Delta\omega\simeq 3.5 {\rm meV}$ and $\Delta\omega\simeq 5 {\rm meV}$ for $B=20 {\rm T}$ and $B = 40 {\rm T}$, respectively.
}
\label{fig:chiral_opt}
\end{figure}

Second, when ${\bf B}$ is neither parallel nor perpendicular to the mirror, all four nodes have different optical gaps and absorption intensities, regardless of ${\bf q}$.
Hence, for a finite frequency window, all the optical absorption takes place at a single Weyl node (see Fig.~\ref{fig:chiral_opt}).
This constitutes a complete {\em chiral} valley polarization and suggests a new way to induce a chiral chemical potential imbalance.

Third, in the non generic case where ${\bf B}$ is perpendicular to the tilt vectors of all four nodes (in our model, this is possible only if 
${\bf t}$ is either parallel or perpendicular to the mirror plane), the valley polarization is completely turned off.
This remark will become relevant in Sec.~\ref{sec:Disc}.

\subsection{Observable signatures of the valley polarization}

The complete, magnetic-field-induced valley polarization found above has topological origin in the sense that it results exclusively from optical transitions involving the chiral LL.
Unlike nonchiral LLs, a chiral LL intersects the Fermi energy only once.
Then, the interband transitions from (or to) the chiral Landau level explore only one side ($k_\parallel>0$ or $k_\parallel<0$) of the anisotropic dispersion relation in the $k_\parallel$ direction.
This peculiarity is responsible for the complete valley polarization.

Such valley polarization is fundamentally different from the one predicted and measured in two dimensional transition metal dichalcogenides (TMDC).\cite{Niu2008, Zeng2012}
In TMDC, RCP photons and LCP photons are absorbed in different valleys, as a consequence of valley-dependent optical selection rules.
The valleys are degenerate in energy and no magnetic field is required in order to obtain a full valley polarization.
Moreover, the valley-dependent selection rules enable a detection of the valley polarization in photoluminescence.\cite{Zeng2012}

In our case, the 100\% valley polarization emerges not from selection rules, but from a magnetic-field-induced lifting of the valley degeneracy and from the unidirectionality of the chiral Landau levels.
In WSM, the complete valley polarization takes place only at certain frequency intervals, and is controllable with the magnetic field and with the polarization (or propagation direction) of the incident light. 
Under pulsed irradiation, the valley polarization in WSM has likely a shorter lifetime than the one in TMDC, because of the gapless energy spectrum.
In addition, because the complete valley polarization in WSM does not originate from optical selection rules, it is not experimentally accessible by photoluminescence.
Next, we propose an alternative experimental signature.

\begin{figure}[t]
\includegraphics[width=0.45\textwidth]{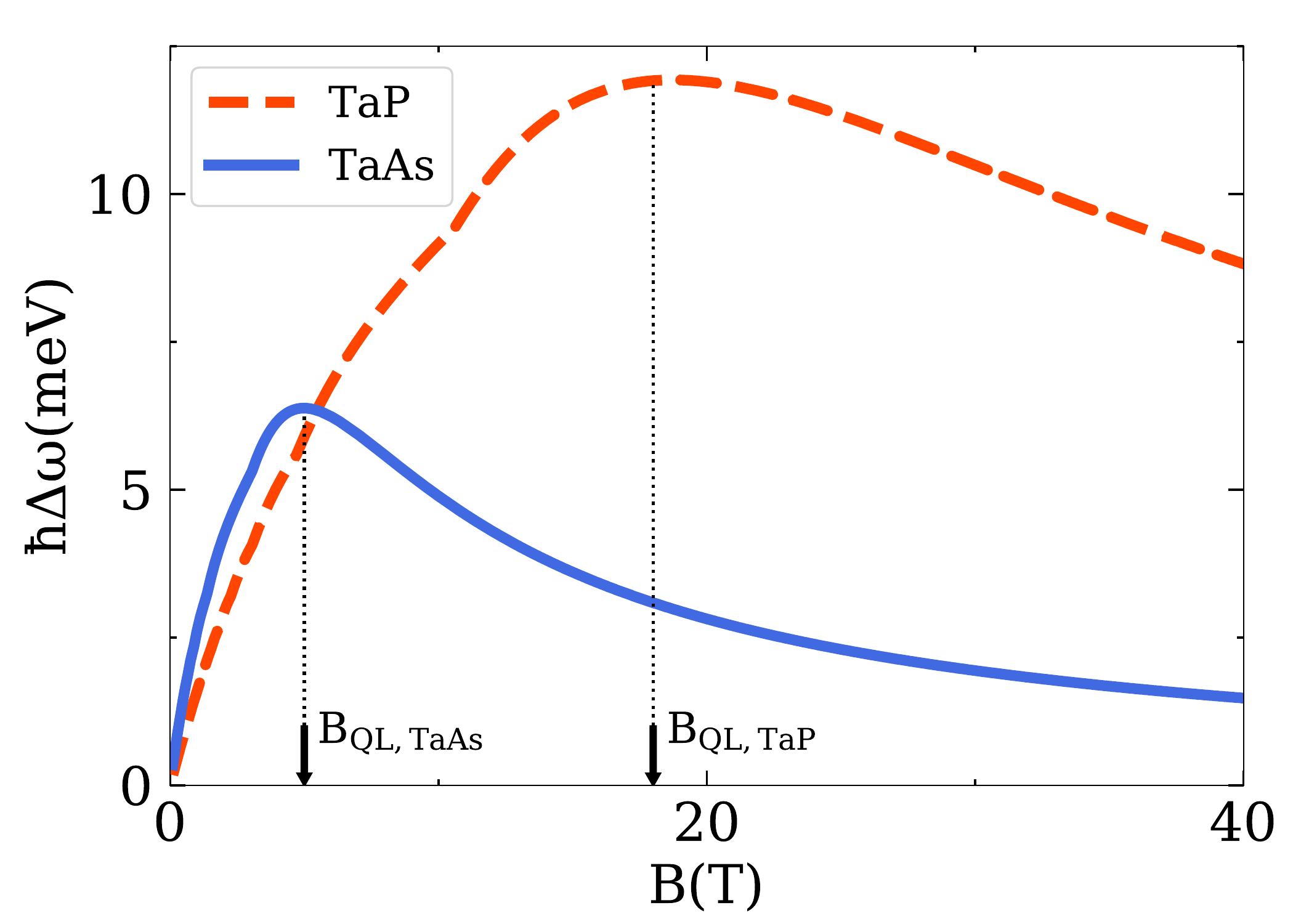}
\caption{
  Frequency-splitting of the $0 \to 1$ (for TaAs) and $-1 \to 0$ (for TaP)  transition taking place in the vicinity of W2 Weyl nodes, as a function of the magnetic field.
  The configuration is ${\bf B}||{\bf q} ||\hat{\bf z}$.
The parameter values are taken from Ref.~[\onlinecite{Grassano2018}].
For TaAs, we have $\mu = 13 \text{meV}$ (zero-field value), $t_\parallel = 0.47 $, $v_F = 3.0 \times 10^5 \text{m/s}$ and $\gamma=1.09$ (see text for definitions). 
For TaP, $\mu = - 21 \text{meV}$ (zero-field value), $t_\parallel = 0.5 $, $v_F = 2.77 \times 10^5 \text{m/s}$ and $\gamma=1.16$.
The arrows indicate the magnetic field value ($B_{\rm QL}$) at which the system enters the quantum limit (for $B>B_{\rm QL}$, only the chiral LL intersects with the chemical potential). 
}
\label{fig:delomega}
\end{figure}

\begin{figure}[t]
\includegraphics[width=0.4\textwidth]{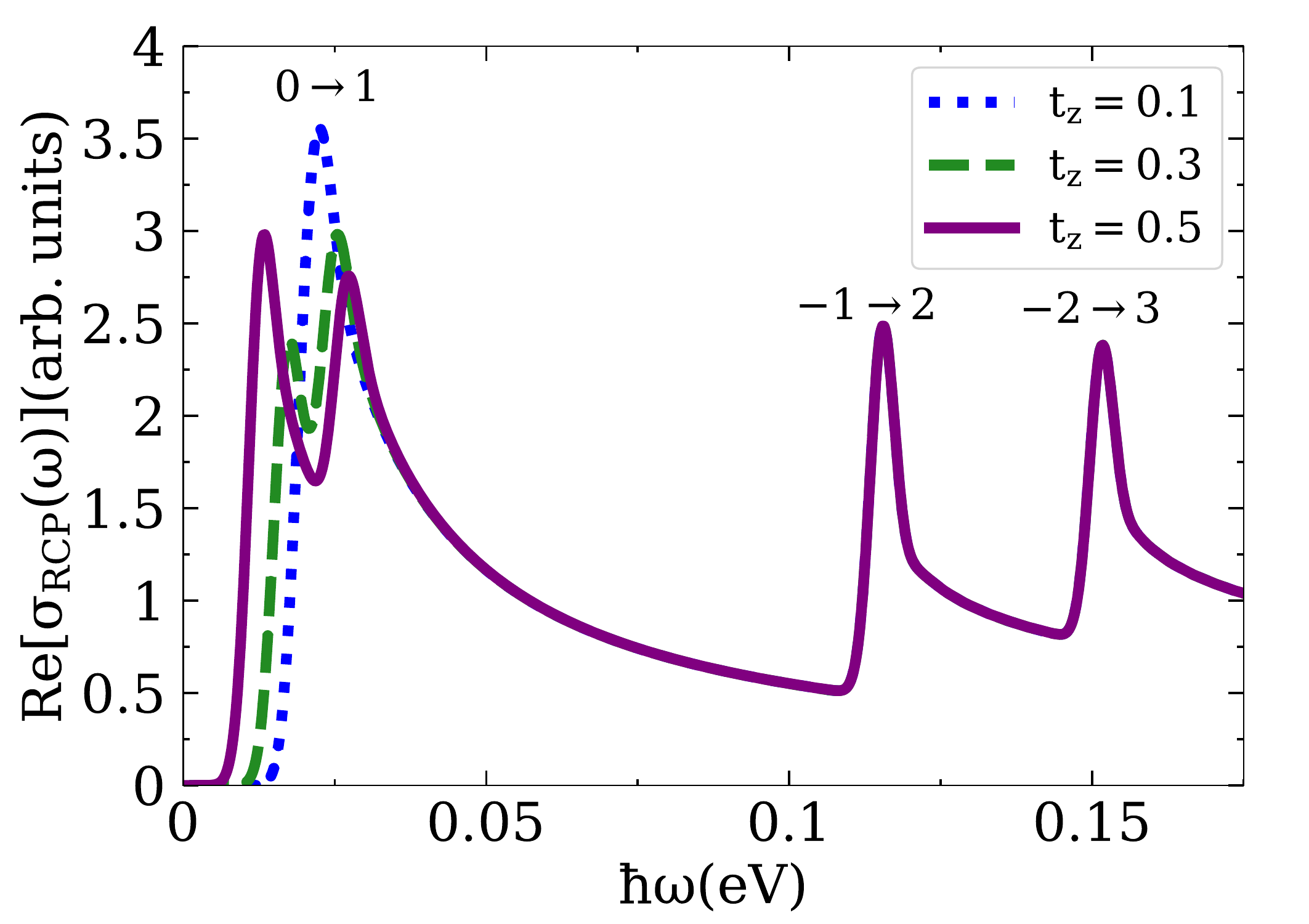}
\includegraphics[width=0.4\textwidth]{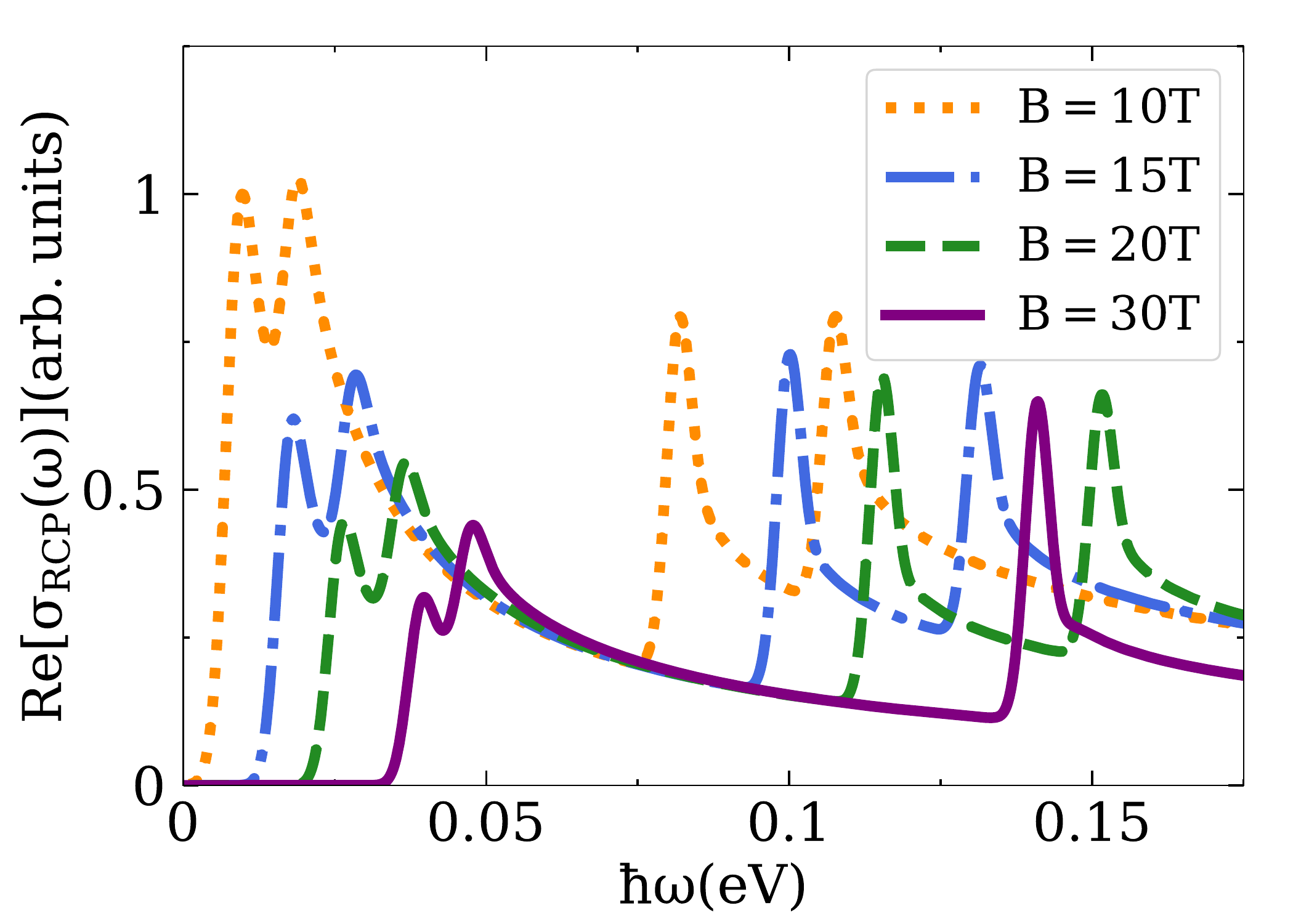}
\includegraphics[width=0.4\textwidth]{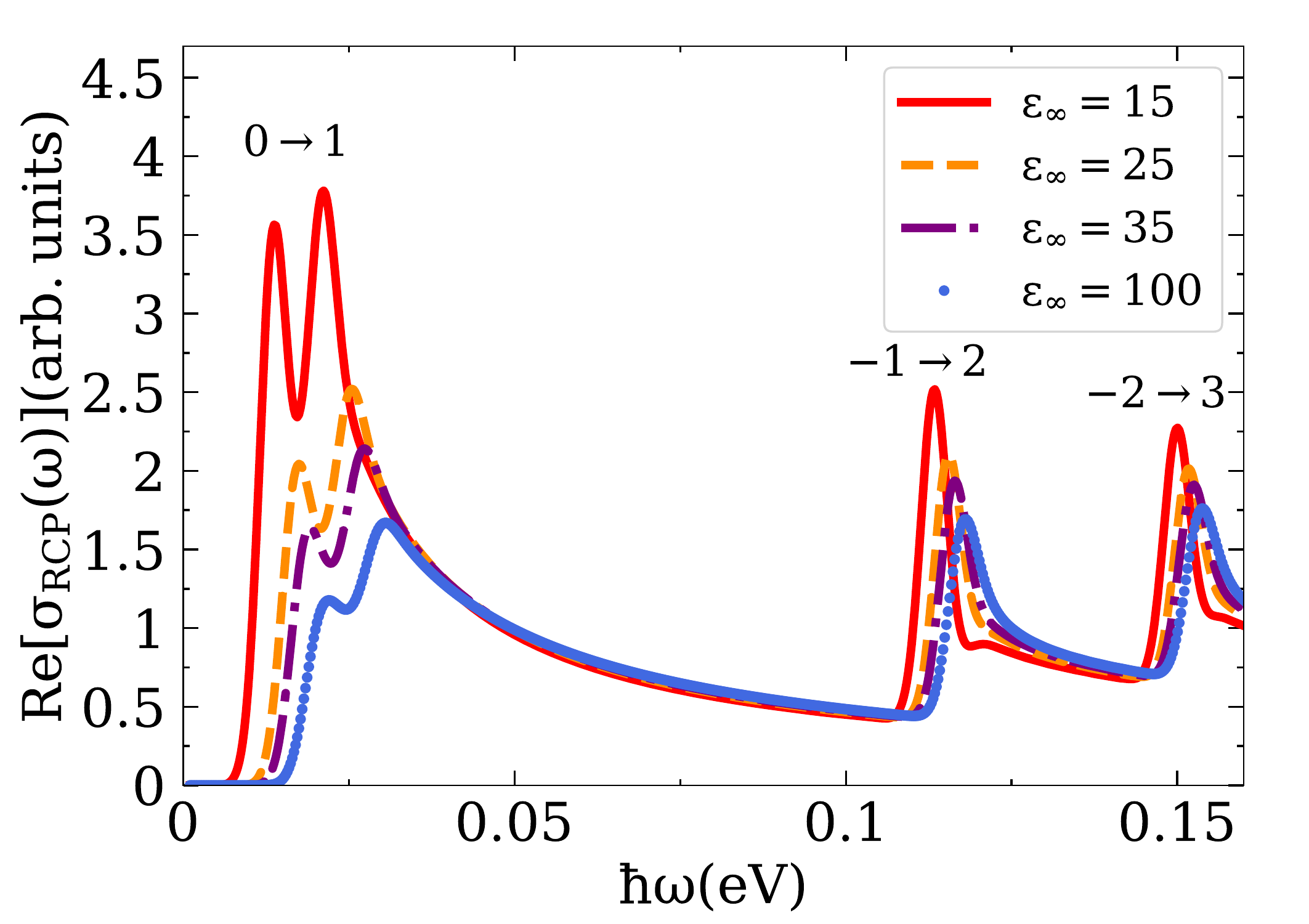}
\caption{
  Total optical absorption in the four-node Weyl semimetal model, with ${\bf B}||{\bf q}||{\bf t}||\hat{\bf z}$ and RCP light.
  Unless otherwise stated, the parameter values are the same as in Fig.~\ref{fig:abs_field}.
  The complete valley polarization (anticipated by Figs. \ref{fig:abs_field} and \ref{fig:valpol_tilt}) manifests through the splitting of the $0 \to 1$ transition.
  The absorption peaks appearing at higher frequencies correspond to the $-1 \to 2$ and $-2 \to 3$ inter Landau level transitions.
  The splitting feature becomes more pronounced as the tilt increases (top panel) or as electron-electron interactions become stronger (bottom panel, with $\mathbf{t} = (0,0,0.3)$).
  The splitting gradually diminishes deep in the quantum limit (middle panel).} 
\label{fig:opttilt}
\end{figure}

The clearest experimental signature of the valley polarization consists of a splitting of $0\to |n|$ (or $-|n|\to 0$) inter Landau-level transitions into two subpeaks (or possibly more in the total absorption spectrum, depending on the direction of ${\bf B}$).
Out of these, the splitting of the $0\to 1$ or the $-1\to 0$ transition is indicative of a complete valley polarization.
For these transitions, the separation in frequency between the split peaks is approximatively equal to the frequency interval $\Delta\omega$ in which the valley polarization is complete.
Unfortunately, the splitting alone cannot tell us which ones of the nodes absorb the light in the frequency interval $\Delta\omega$.

In the absence of Coulomb interaction, it is straightforward to determine analytically the dependence of $\Delta\omega$ on the magnetic field, tilt and the chemical potential.
For concreteness, we will consider $\mu>0$ and an RCP light in the direction parallel to the magnetic field, the latter being perpendicular to the mirror plane.
In this case, the complete valley polarization between nodes $(1,2)$ and $(3,4)$ emerges from $0\to 1$ optical transitions.
Using Eq.~(\ref{eq:tchouma}), we arrive at
\begin{align}
  \hbar \Delta \omega = \Bigg\lvert &- \frac{2\mu \gamma t_\parallel}{1-\gamma^2 t_\parallel^2}+\sqrt{\frac{2\hbar^2 v_F^2}{\gamma^3 \ell_B^2} + \frac{\mu^2}{(1 - \gamma t_\parallel)^2}}\nonumber\\
  & - \sqrt{\frac{2\hbar^2 v_F^2}{\gamma^3 \ell_B^2} + \frac{\mu^2}{(1 + \gamma t_\parallel)^2}} \, \Bigg\lvert. \label{eq:delomegafinal}
\end{align}
Equation (\ref{eq:delomegafinal}) describes the difference in optical gaps between time-reversed partner nodes.
It shows that $\Delta\omega =0$ when either $\mu$, $t_\parallel$ or $B$ vanish.
At high magnetic fields, where $\ell_B^{-1} \gg \mu/(\hbar v_F)$, Eq.~(\ref{eq:delomegafinal}) leads to $\hbar \Delta \omega = 2 |\mu \gamma t_\parallel/( 1 - \gamma^2 t_\parallel^2)| $.
In this regime, $\Delta\omega$ depends on the magnetic field indirectly through the chemical potential.
We calculate the field-dependence of the chemical potential by demanding that the density of carriers (number of electrons at positive energy minus the number of
holes at negative energy) remain invariant as the field is varied.
In the quantum limit, as the field is made stronger, the chemical potential decreases towards zero and hence so does $\Delta\omega$.
In the weak magnetic field regime, where $\ell_B^{-1} \ll \mu/(\hbar v_F)$, Eq.~(\ref{eq:delomegafinal})  leads to $\Delta \omega = 2 e B v_F^2 t_\parallel/(\gamma^2 \mu)$.
In this case, the dependence of the chemical potential on the magnetic field is weak and $\Delta\omega$ scales approximately linearly with $B$, as well as with $t_\parallel$.
The dependence of $\Delta\omega$ for the entire range of magnetic fields is shown in Fig. \ref{fig:delomega}.
The maximum value of $\Delta \omega$ is attained when $\mu\simeq \hbar v_F/l_B$, i.e. when the chemical potential intersects only the chiral LL but lies close to the first nonchiral LL.

Figure \ref{fig:opttilt} shows the peak splitting in the total optical absorption for the case in which ${\bf B}$ and ${\bf q}$ are aligned with one another and are perpendicular to the mirror plane.
As anticipated by Eq.~(\ref{eq:delomegafinal}), the splitting grows monotonically with $t_\parallel$.
The results do not change qualitatively in the presence of Coulomb interactions.
On a quantitative level, electron-electron interactions help observe the double peak feature by making the absorption peaks sharper (see the bottom panel of Fig. (\ref{fig:opttilt})).
Because $\Delta\omega$ is about $10 \, {\rm meV}$ for reasonable tilt and magnetic fields, the double-peak feature in the optical conductivity should be detectable for a typical experimental resolution.\cite{XuDai2016,Neubauer2018, Yuan2018, Jiang2018}

\section{Application to real Weyl semimetals}\label{sec:Disc}






Thus far, we have used a toy model of four Weyl nodes to predict an optical valley polarization that has topological origin and takes place in a magnetic field.
The measurable consequence of this polarization is a splitting of the inter Landau level transitions $0\to |n|$ and $-|n|\to 0$ in the optical absorption spectrum.

The toy model has been useful to elucidate the symmetry requirements for the generation and control of the valley polarization.
In this section, we go further by extrapolating our results to real WSM with broken inversion symmetry, with a focus on TaAs, TaP, NbAs and NbP.
\cite{Weng2015, ChiChengLee2015, Grassano2018}
Among these, TaAs and TaP are the most promising for our purposes because the approximation of linear energy dispersion is valid in a wider frequency range in these materials.

TaAs and related compounds contain two symmetry-inequivalent sets of Weyl nodes, denoted W1 and W2, with multiplicities of 8 and 16 (respectively).
The eight symmetry-equivalent W1 nodes are located at the $k_z=0$ plane, which is invariant under time-reversal.
This fact, combined with the $C_{4v}$ symmetry of these crystals, precludes any tilt of the W1 nodes in the $z$ direction.
The sixteen symmetry-equivalent W2 nodes are located at two planes of constant but nonzero value of $k_z$.
Thus, the Weyl cones at W2 nodes are allowed to have (and they do have) nonzero tilts along the $z$ direction.

The difference in tilts between the W1 and W2 nodes leads to a significant disparity in their optical absorption spectra, which manifests itself most simply when the magnetic field is applied along the $z$ direction.
When ${\bf B}||{\bf q}||\hat{\bf z}$, all the mirror symmetries (${\cal M}$) as well as the time-reversal symmetry (${\cal T}$) are broken, but their product ${\cal M}{\cal T}$ and the fourfold rotation symmetry axis ($C_4$) are preserved.
Because the eight W1 nodes can be related to one another by either ${\cal M}{\cal T}$ or $C_4$, they all display an identical optical absorption.
In other words, there is no valley polarization and no splitting of the inter LL transitions for W1 nodes. 
In contrast, ${\cal M}{\cal T}$ and $C_4$ connect only eight of the sixteen W2 nodes.
The symmetries relating the two octuplets being broken by the field, a valley polarization and a splitting of the inter LL transitions ensue at W2 nodes. 

\begin{figure}[t]
\includegraphics[width=0.45\textwidth]{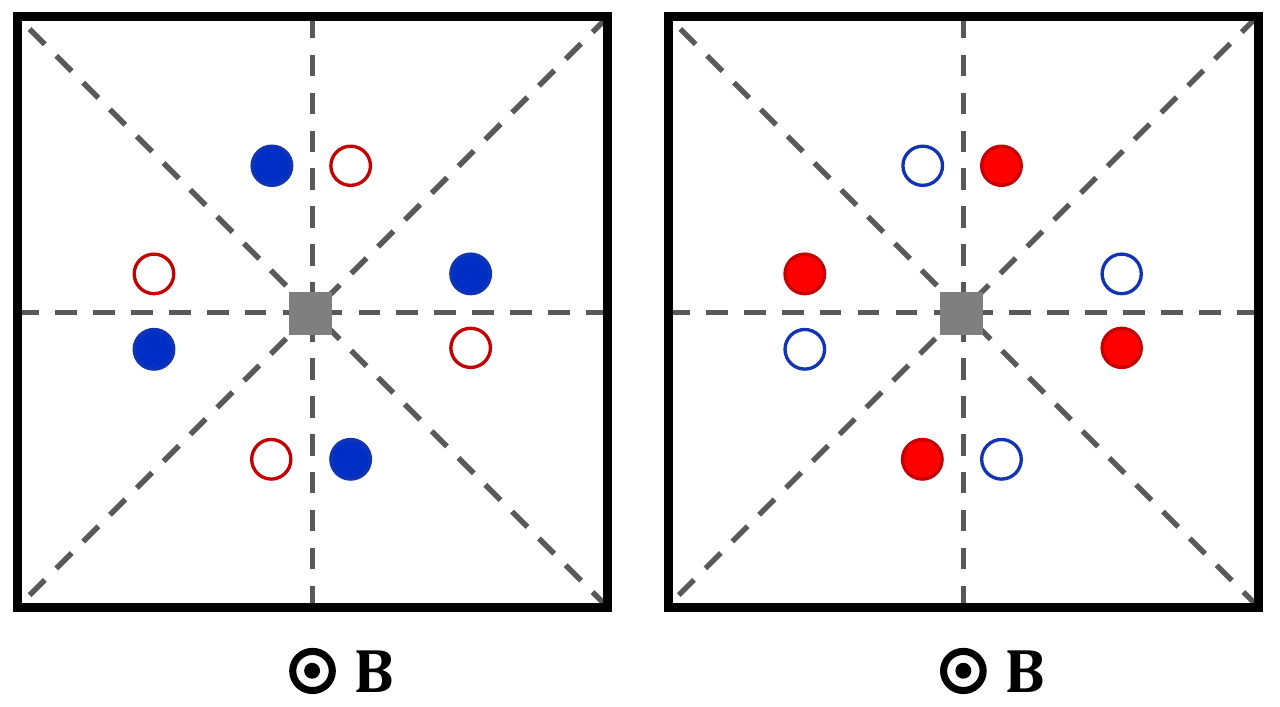}
\caption{Cartoon of the optical absorption intensity for W2 Weyl nodes in TaAs. The magnetic field and the propagation direction of the incident light are along the $z$ direction.
  The grey square denotes the $C_4$ axis, while the dashed lines denote mirror planes.  The left (right) panel shows eight Weyl nodes located in the constant $k_z>0$ ($k_z<0$) plane.
  The solid (empty) circles denote Weyl nodes that absorb (do not absorb) light for a finite frequency range of the incident light. The red and blue colors denote opposite chiralities.}
\label{fig:TaAs_cartoon}
\end{figure}

The pattern of the absorption intensity in momentum space  is illustrated in Fig.~\ref{fig:TaAs_cartoon}.
Interestingly, the absorption is chiral for the Weyl nodes located at the same plane of constant $k_z$.
Although the net optically induced chirality is zero when summing over the two planes of opposite $k_z$, the dipole moment of the chirality is not. 

Another interesting possibility, alluded to in Sec.~\ref{sec:res1}, is the emergence of a 100 \% {\em chiral} valley polarization when the magnetic field is applied along a direction without any particular symmetry.
This finding anticipates large and magnetically tunable photogalvanic effects\cite{Golub2017, Golub2018, Kharzeev2018} in WSM.

It is important to reiterate that the valley polarization discussed in this work occurs between Weyl nodes that are symmetry-equivalent in the absence of the field.
There is, of course, a complete valley polarization {\em between} W1 and W2 nodes for a significant frequency interval of the incident light even at zero field, because these nodes are not equivalent by symmetry and have therefore different optical gaps.
However, this ``trivial'' valley polarization is not interesting because it cannot be controlled (e.g. reversed) with external perturbations.

We conclude this section by commenting on a number of recent Landau level spectroscopy measurements in WSM.\cite{Jiang2018, Yuan2018}
Reference [\onlinecite{Jiang2018}] has reported splitting features in the optical transitions involving the chiral Landau levels of W2 nodes in NbP, which are reminiscent to the ones predicted in this work. 
In NbP, the separation between Weyl nodes is small and thus the authors use a model of two coupled nodes to interpret their data.
In contrast, our model assumes isolated Weyl nodes.

In Ref. [\onlinecite{Jiang2018}], the splitting feature is attributed to a difference in the optical gaps between $0\to 1$ and $-1\to 0$ transitions, which are activated simultaneously by the linearly polarized light that the authors shine on the sample.  
This theoretical interpretation requires adding a Zeeman term to the authors' model Hamiltonian, with an {\em ad hoc} value of the $g$ factor.
According to the authors, the Zeeman term models the breaking of particle-hole symmetry by inducing an asymmetry between the conduction and valence band Landau levels.
It is not clear in their discussion why the particle-hole symmetry breaking does not visibly manifest in W1 nodes, where no splitting feature is observed.

In our theory, the $0\to 1$ and $-1\to 0$ transitions, which are separately excited by a circularly polarized light,
 can {\em each} split in two.
The splitting results from the breaking of an effective antiunitary symmetry of the Weyl Hamiltonian by the tilt of the cones or by nonlinear terms in the energy dispersion, as well as from the breaking of time reversal symmetry by a magnetic field.
No Zeeman terms are required to explain the splitting, and the breaking of particle-hole symmetry is not necessary.
In fact, App. \ref{App_nonlin} shows that a particle-hole symmetric model with nonlinear terms in the energy dispersion leads to a splitting of the $0\to 1$ transition.
Moreover, as mentioned above, our theory offers a possible explanation on why the splitting is seen at interband transitions taking place near the W2 nodes, but not in those occurring in the vicinity of W1 nodes.

\section{Summary and conclusions}
\label{sec:conc}


In summary, we have presented a theory of the interband magneto-optical absorption in a Weyl semimetal with broken inversion symmetry.
We have considered generic orientations of the magnetic field and the light propagation, and have included the effect of Coulomb interactions approximately.
All of our calculations have been carried out for a minimal model in which the coupling between different Weyl nodes is neglected.
This limits the validity of our results to Weyl semimetals in which the internode separation exceeds largely the inverse of the magnetic length.
Similarly, we have assumed that the pseudospin of the Weyl fermions transforms in the same way as the spin under point group symmetry operations.
Although this is not true in general, we have been careful in ensuring that our main results are qualitatively applicable to real Weyl semimetals.

The main prediction of our work is the existence of a complete valley polarization in strong magnetic fields.
Since the application of a magnetic field reduces the symmetry of the crystal, the optical absorption intensity differs at Weyl nodes that were equivalent by symmetry at zero field. 
For a sizeable frequency interval $\Delta\omega$ of the incident light, the difference in the absorption intensity reaches 100\%.
This complete valley polarization has topological origin because it results exclusively from interband transitions involving the chiral Landau level.
In principle it is possible to control which Weyl nodes will absorb the light by an appropriate choice of the directions of the magnetic field and light propagation.
Remarkably, when the magnetic field points along a low-symmetry direction, the entire optical absorption can be confined to the vicinity of a single Weyl node. 

There are various requirements for the occurrence of the complete valley polarization.
First, if the Hamiltonian of a Weyl node $\tau$ is given by $h_\tau({\bf k})$, it is necessary that $h_\tau({\bf k}) \neq \Theta^{-1} h_\tau(-{\bf k}) \Theta$, where $\Theta=i \sigma^y K$ is an antinuitary operator (different from time reversal or charge conjugation),  ${\bf k}$ is the momentum measured from the Weyl node, $\sigma^y$ is the $y-$component of a Weyl fermion's pseudospin and $K$ is the complex conjugation operator.
The tilting of Weyl cones or the presence of nonlinear terms in the energy spectrum are necessary in order to ensure this condition.
Second, the Fermi energy must be located away from the charge neutrality point.
The valley polarization is easiest to observe at the onset of the quantum limit, i.e. when the chemical potential intersects only the chiral Landau level but is close to the first nonchiral Landau level.
Third, in a Weyl semimetal with tilted cones, the direction of the magnetic field must have a nonzero projection along the direction of the tilt.

In the TaAs family of Weyl semimetals, the third condition above is not satisfied for the W1 Weyl nodes when the magnetic field is aligned with the $c$ axis.
Thus, we expect a negligible valley polarization for the W1 nodes in this configuration.
\footnote{There could still be a nonzero effect coming from nonlinear terms in the energy spectrum, though these are unimportant if the Fermi energy is close to the Weyl nodes.}
In contrast, all three conditions are realized for the W2 Weyl nodes, which should therefore host a complete valley polarization.
This difference suggests a way to optically distinguish W1 and W2 nodes.

The simplest experimental signature of the complete valley polarization is a splitting of the inter Landau level transitions involving the chiral Landau level (in particular, the $0\to 1$ or $-1\to 0$ transitions).
Coulomb interactions accentuate this splitting through an increase of the optical spectral weight in the vicinity of the absorption peaks.
The magnitude of the splitting roughly coincides with $\Delta\omega$, which attains a maximum at the onset of the quantum limit and gradually decays as the field is made stronger (because the chemical potential approaches the charge neutrality point).
The precise value of $\Delta\omega$ depends on the density of carriers, the magnetic field and the tilt vector (or the coefficients in front of the nonlinear terms in the energy dispersion); it can reach $\simeq 10\, {\rm meV}$ for realistic band parameters and attainable magnetic fields.

The present study suggests various avenues for future research.
For example, it would be interesting to understand the dynamics and steady states of the valley polarization.
Similarly, it would be of interest to establish transport-based diagnostic tools for the complete valley polarization, so that one can know experimentally which nodes are absorbing photons and which ones are not. 
Finally, as mentioned in Secs. \ref{sec:intro} and \ref{sec:Disc}, a Landau level splitting reminiscent to the one predicted in this work has been recently observed in experiment, though the theoretical interpretation offered therein differs from ours.
More experimental and theoretical work is needed to settle this issue. 

\section*{ACKNOWLEDGEMENTS}

We acknowledge financial support from Qu{\'e}bec RQMP, Canada's NSERC, and the Canada First Research Excellence Fund. The numerical calculations were performed on computers provided by Calcul Qu{\'e}bec and Compute Canada. We thank P.~Rinkel, S.~Acheche, X.~Yuan and B.~Ramshaw for helpful discussions.

\appendix

\section{Optical conductivity in the generalized random-phase approximation}\label{App_GRPA}

In this appendix, we present an approximate method that allows to calculate the optical conductivity of an interacting Weyl semimetal in a magnetic field.

Our starting point is Eq. (\ref{eq:optcond}).
Explicitly, the real-time retarded current response function reads
\begin{equation}
\chi_{J_{\alpha }J_{\beta }}^{R}\left( \mathbf{q},t\right) =-\frac{i}{%
{\cal V}\hbar }\left\langle \left[ J_{\alpha }\left( \mathbf{q},t\right)
,J_{\beta }\left( -\mathbf{q},0\right) \right] \right\rangle \Theta \left(
t\right) , \label{grpa23}
\end{equation}
where ${\cal V}$ is the volume of the system, $\Theta \left( t\right) $ the Heaviside
step function and $J_\alpha$ is the $\alpha$-th component of the current operator $\mathbf{J}$.
Next, we derive an equation for $\chi _{J_{\alpha }J_{\beta }}^{R}$ within the generalized random-phase approximation (GRPA).\citep{BaymKadanoff1961, Baym1962}

We start with the interacting Hamiltonian
\begin{align}
&{\cal H}-\mu N = \sum_{a}\int d \mathbf{u}\, \Psi _{a}^{\dagger }\left( \mathbf{u}\right) %
\left[ h_{a}\left( \mathbf{u}, i\partial_{\bf u}\right) -\mu\right] \Psi _{a}\left( \mathbf{u}\right) \label{grpa22} \\
&+\frac{1}{2}\sum_{a,b}\int d\mathbf{u}\, d\mathbf{u'}\,\Psi _{a}^{\dagger
}\left( \mathbf{u}\right) \Psi _{b}^{\dagger }\left( \mathbf{u}^{\prime
}\right) V\left( \mathbf{u}-\mathbf{u}^{\prime }\right) \Psi _{b}\left( 
\mathbf{u}^{\prime }\right) \Psi _{a}\left( \mathbf{u}\right),  \nonumber
\end{align}
where 
\begin{equation}
  \Psi _{a}\left( \mathbf{u}\right) =\left(\begin{array}{c} \Psi_{a\uparrow}({\bf u}) \\ \Psi_{a\downarrow}({\bf u})\end{array}\right) =\sum_{p}w_{a,p}\left( \mathbf{u}\right) c_{a,p} 
\end{equation}%
is an operator that destroys an electron at a position $\mathbf{u}\equiv\left( \mathbf{r},z\right)$ and at Weyl node $a$,
${\bf r}$ is the position of the electron in the plane perpendicular to the magnetic field (${\bf B}||\hat{\bf z}$ in this Appendix), the $\uparrow$ and $\downarrow$ subscripts stand for up and down pseudospin components of the spinor, $p$ stands for the collection of good quantum numbers characterizing an eigenstate,  and $w_{a,p}\left( \mathbf{u}\right)$ is a two-component eigenspinor of the non-interacting $2\times 2$ Hamiltonian $h_{a}$.
The operator $c_{a,p}\left( c_{a,p}^{\dag }\right) $ annihilates (creates) an electron in a state $p$ at a node $a$.

In Eq.~(\ref{grpa22}), we considered a Coulomb interaction that is diagonal in the node and pseudospin spaces, and defined
\begin{equation}
V\left( \mathbf{u}-\mathbf{u}^{\prime }\right) =\frac{e^{2}}{4\pi \varepsilon_{\infty} \left\vert \mathbf{u}-\mathbf{u}^{\prime
}\right\vert }.
\end{equation}%
In the interacting term of Eq.~(\ref{grpa22}), a scalar product between spinors belonging to the same position index ${\bf u}$ or ${\bf u}'$ is implied.

As usual, the retarded current response function is obtained by analytical continuation of the 
two-particle Matsubara Green's function
\begin{equation}
  \label{eq:jmatsu}
\chi _{J_{\alpha }J_{\beta }}\left( \mathbf{q},\tau \right) =-\frac{1}{%
{\cal V} \hbar }\left\langle T_{\tau }J_{\alpha }\left( \mathbf{q},\tau
\right) J_{\beta }\left( -\mathbf{q},0\right) \right\rangle ,
\end{equation}%
where $\tau $ is the imaginary time and $T_{\tau }$ is the imaginary time
ordering operator. The Fourier transform of this function is
\begin{equation}
\chi _{J_{\alpha }J_{\beta }}\left( \mathbf{q},i\Omega _{n}\right)
=\int_{0}^{\beta \hbar }d\tau e^{i\Omega _{n}\tau }\chi _{J_{\alpha
}J_{\beta }}\left( \mathbf{q},\tau \right) ,
\end{equation}%
where $\Omega _{n}=2n\pi /\beta \hbar $ is a bosonic Matsubara frequency and $\beta =1/k_{B}T$, with $k_{B}$ the Boltzmann constant.
The retarded response giving the optical conductivity in Eq.~(\ref{eq:optcond}) is obtained by performing an analytic continuation ($i\Omega_{n}\rightarrow \omega +i\delta $) of $\chi_{J_\alpha J_\beta}({\bf q},i\Omega_n)$.
At the end of the calculation, we will take the zero temperature limit.

After performing the minimal substitution in $h_{a}\left( \mathbf{u}, i\partial_{\bf u}\right)$, the current operator in first quantization is given by
\begin{equation}
j_{a,\alpha }=-\frac{\delta h_{a}}{\delta A_{\alpha }}, \label{grpa3}
\end{equation}%
where $A_\alpha$ is a component of the vector potential.
The operator $j_{a, \alpha}$ can be represented as a $2\times 2$ matrix in pseudospin space.
In second quantization, the Fourier transform of the current operator is 
\begin{eqnarray}
J_{a,\alpha }\left( \mathbf{q}\right)  &=&\int d\mathbf{u}\,\Psi _{a}^{\dag
}\left( \mathbf{u}\right) e^{-i\mathbf{q}\cdot \mathbf{u}}j_{a,\alpha }\Psi
_{a}\left( \mathbf{u}\right)   \label{grpa1} \\
&=&\sum_{p_{1},p_{2}}\Lambda _{p_{1},p_{2}}^{\left( a,\alpha \right) }\left( 
\mathbf{q}\right) c_{a,p_{1}}^{\dag }c_{a,p_{2}} , \nonumber
\end{eqnarray}%
where the time dependence is implicit in the creation and annihilation operators.
The matrix elements in Eq.~(\ref{grpa1}) are
\begin{equation}
  \label{eq:mel}
\Lambda _{p_{1},p_{2}}^{\left( a,\alpha \right) }\left( \mathbf{q}\right)
=\int d\mathbf{u}\,e^{-i\mathbf{q}%
\cdot \mathbf{u}} \left[w_{a,p_{1}}^{\dag }\left( \mathbf{u}\right) \cdot j_{a,\alpha } \cdot w_{a,p_{2}}\left( \mathbf{u}\right)\right] .
\end{equation}%
In Eq.~(\ref{eq:mel}), we have explicitly indicated the scalar products.
Inserting Eq.~(\ref{grpa1}) in Eq.~(\ref{eq:jmatsu}) and using $J_\alpha = \sum_a J_{a,\alpha}$, we have
\begin{align}
  \chi _{J_{\alpha }J_{\beta }}\left( \mathbf{q},\tau\right) &=%
\frac{1}{{\cal V}\hbar }\sum_{a,b}\sum_{p_{1}, p_2, p_3, p_{4}}\Lambda
_{p_{1},p_{2}}^{\left( a,\alpha \right) }\left( \mathbf{q}\right) \Lambda
_{p_{3},p_{4}}^{\left( b,\beta \right) }\left( -\mathbf{q}\right) \nonumber \\
&\chi_{p_{1},p_2, p_3,p_{4}}^{\left( a,a,b,b\right) }\left( \tau\right) , \label{grpa7}
\end{align}%
where we have defined the two-particle Green's function
\begin{align}
\chi _{p_{1}, p_2, p_3,p_{4}}^{\left( a,b,c,d\right) }&\left( \tau _{1}-\tau
_{2}\right) = \\
&-\left\langle T_{\tau }c_{a,p_{1}}^{\dag }\left( \tau
_{1}\right) c_{b,p_{2}}\left( \tau _{1}\right) c_{c,p_{3}}^{\dag }\left(
\tau _{2}\right) c_{d,p_{4}}\left( \tau _{2}\right) \right\rangle . \nonumber
\end{align}

The next task is to derive an explicit expression for $\chi^{(a...d)}_{p_1...p_4}$. 
We begin by recalling the definitions of the single- and two-particle Green's functions, respectively:
\begin{equation}
G_{a,b}\left( 1,2\right) =-\left\langle T_{\tau }\Psi _{a}\left( 1\right)
\Psi _{b}^{\dag }\left( 2\right)\right\rangle 
\end{equation}%
and
\begin{align}
L_{a,b,c,d}\left( 1,2,3,4\right) &=-\left\langle T_{\tau }\Psi _{a}^{\dagger
}\left( 1\right) \Psi _{b}\left( 2\right) \Psi _{c}^{\dagger }\left(
3\right) \Psi _{d}\left( 4\right) \right\rangle \nonumber \\
&+G_{b,a}\left( 2,1\right)G_{d,c}\left( 4,3\right).
\end{align}%
In these expressions, the number $n=\left( \mathbf{u}_{n},\tau _{n}\right) $ combines the position and imaginary time indices.
In GRPA, the single-particle Green's function $G_{a,b}\left( 1,2\right)$ is evaluated in the Hartree-Fock approximation, i.e.
\begin{align}
G_{a,b}\left( 1,2\right) &=G_{a,b}^{0}\left( 1,2\right) \nonumber \\
&+\sum_{c,d}G_{a,c}^{0}\left( 1,\overline{3}\right) \Sigma^{HF}_{c,d}\left( 
\overline{3},\overline{4}\right) G_{d,b}\left( \overline{4},2\right),
\label{grpa4}
\end{align}%
where $G_{a,b}^{0}\left( 1,2\right) $ is the non-interacting Green's function,
\begin{align}
  \label{eq:sigmaHF}
\Sigma _{a,b}^{HF}\left( 1,2\right) &=\frac{1}{\hbar }\delta _{a,b}\delta
\left( 1-2\right) V\left( 1-\overline{3}\right) \sum_{c}G_{c,c}\left( 
\overline{3},\overline{3}^{+}\right) \nonumber \\
&-\frac{1}{\hbar }V\left( 1-2\right)
G_{a,b}\left( 1,2\right) 
\end{align}
is the Hartree-Fock self-energy and $V\left(1-2\right) =V\left( \mathbf{u}_{1}-\mathbf{u}_{2}\right) \delta\left(\tau_{1}-\tau_{2}\right)$.
In Eqs.~(\ref{grpa4}) and (\ref{eq:sigmaHF}), repeated numbers with a bar imply an integral over time and space coordinates.
The superscripts $+$ and $-$ denote times that differ by an infinitesimal amount.

The GRPA for the two-particle Green's function is obtained from a functional derivative of the Hartree-Fock self-energy.\citep{BaymKadanoff1961}
This results in
\begin{widetext}
\begin{eqnarray}
&&L_{a,b,c,d}\left( \mathbf{u}_{1},\mathbf{u}_{1},\mathbf{u}_{2},\mathbf{u}%
  _{2};\tau _{1}-\tau _{2}\right)\nonumber\\
  &&= G_{b,c}\left( \mathbf{u}_{1},\mathbf{u}_{2};\tau _{1}-\tau _{2}\right)
G_{d,a}\left( \mathbf{u}_{2},\mathbf{u}_{1};\tau _{2}-\tau _{1}\right)  
\label{grpa5} \\
&&+\frac{1}{\hbar }\sum_{e, f}\int d\mathbf{u}_{3}\int d\mathbf{u}_{4}\int
d\tau _{3}\,G_{b,e}\left( \mathbf{u}_{1},\mathbf{u}_{3};\tau _{1}-\tau
_{3}\right) G_{e,a}\left( \mathbf{u}_{3},\mathbf{u}_{1};\tau _{3}-\tau
_{1}\right) V\left( \mathbf{u}_{3}-\mathbf{u}_{4}\right) L_{f,f,c,d}\left( 
\mathbf{u}_{4},\mathbf{u}_{4},\mathbf{u}_{2},\mathbf{u}_{2};\tau _{3}-\tau
_{2}\right)   \nonumber \\
&&-\frac{1}{\hbar }\sum_{e,f}\int d\mathbf{u}_{3}\int d\mathbf{u}_{4}\int
d\tau _{3}\,G_{b,e}\left( \mathbf{u}_{1},\mathbf{u}_{3};\tau _{1}-\tau
_{3}\right) G_{f,a}\left( \mathbf{u}_{4},\mathbf{u}_{1};\tau _{3}-\tau
_{1}\right) V\left( \mathbf{u}_{3}-\mathbf{u}_{4}\right) L_{f,e,c,d}\left( 
\mathbf{u}_{4},\mathbf{u}_{3},\mathbf{u}_{2},\mathbf{u}_{2};\tau _{3}-\tau
_{2}\right) . \nonumber
\end{eqnarray}%
Here, we have written an equation for $L$ having only two position and two time coordinates, as this is the quantity needed for the optical conductivity.
Moreover, because we are interested in linear response to an external perturbation, $L$ depends only on the difference between the times $\tau_1$ and $\tau_2$. 
The first term in Eq. (\ref{grpa5}) is the Hartree-Fock response.
The second and third lines capture the polarization and excitonic corrections, respectively.
Note that the excitonic corrections involve a three-point Green's function (in space), in contrast with the other terms.

We can get $\chi _{p_{1}, p_2, p_3, p_{4}}^{\left( a,b,c,d\right) }\left( i\Omega
_{n}\right) ,$ the function needed in Eq. (\ref{grpa7}), by expanding $L_{a,b,c,d}$ and $G_{a,b}$ in Eq. (\ref{grpa5}) on the eigenspinor basis.
For example,
\begin{align}
  \label{eq:L}
L_{a,b,c,d}\left( \mathbf{u}_{1},\mathbf{u}_{1},\mathbf{u}_{2},\mathbf{u}%
_{2};\tau _{1}-\tau _{2}\right) &= \sum_{p_{1} \cdots p_{4}}
\left[w_{a,p_{1}}^{\dag}\left( \mathbf{u}_{1}\right) \cdot w_{b,p_{2}}\left( \mathbf{u}_{1}\right)\right]
\left[w_{c,p_{3}}^{\dag }\left( \mathbf{u}_{2}\right) \cdot w_{d,p_{4}}\left( \mathbf{u}%
_{2}\right)\right] \chi _{p_{1} \cdots p_{4}}^{\left( a,b,c,d\right) }\left( \tau
_{1}-\tau _{2}\right)  
\end{align}
and
\begin{align}
G_{a,b}\left( \mathbf{u}_{1},\mathbf{u}_{2};\tau \right)
&=\sum_{p_{1},p_{2}}w_{a,p_{1}}\left( \mathbf{u}_{1}\right) w_{b,p_{2}}^{\dag
}\left( \mathbf{u}_{2}\right) G_{p_{1},p_{2}}^{\left( a,b\right) }\left(
\tau \right),
\end{align}
with the definition
\begin{equation}
G_{p_{1},p_{2}}^{\left( a,b\right) }\left( \tau\right) =-\left\langle
T_{\tau }c_{a,p_{1}}\left( \tau\right) c_{b,p_{2}}^{\dag }\left(
0\right) \right\rangle .
\end{equation}

Since we ultimately want to compute the current-current response defined in Eq.~(\ref{grpa7}), the scalar product of the spinors must be taken as indicated in Eq.~(\ref{eq:L}).
Moreover, the spinors associated with the single-particle Green's functions in Eq.~(\ref{grpa5}) must be treated according to the rule defined for the Coulomb interaction, namely, the scalar product is to be taken between spinors with the same position ${\bf u}_n$.
For example, in the last line of Eq.~(\ref{grpa5}), the product of the spinors must be interpreted as
\begin{equation}
  \left[w^\dagger_a({\bf u}_1) \cdot w_b({\bf u}_1)\right]   \left[w^\dagger_c({\bf u}_2) \cdot w_d({\bf u}_2)\right]   \left[w^\dagger_e({\bf u}_3) \cdot w_e({\bf u}_3)\right]   \left[w^\dagger_f({\bf u}_4) \cdot w_f({\bf u}_4)\right],
\end{equation}
with the indices $p$ left implicit.
Likewise, in the second line of Eq.~(\ref{grpa5}), we have $  \left[w^\dagger_a({\bf u}_1) \cdot w_b({\bf u}_1)\right]   \left[w^\dagger_c({\bf u}_2) \cdot w_d({\bf u}_2)\right]$ for the spinors associated with the product of the two single-particle Green's functions.

If we strip all terms in Eq. (\ref{grpa5}) of their eigenspinor factors, we get
\begin{eqnarray}
\chi _{p_{1}, p_2, p_3, p_{4}}^{\left( a,b,c,d\right) }\left( i\Omega _{n}\right) 
&&=\chi _{p_{1}, p_2, p_3, p_{4}}^{\left( 0\right) \left( a,b,c,d\right) }\left(
i\Omega _{n}\right)   \label{grpa6} \\
&&+\frac{1}{\hbar}\frac{1}{{\cal V}}\sum_{e,f}\sum_{p_{5},p_{6},p_{7},p_{8}}%
\sum_{\mathbf{q}}\chi _{p_{1},p_{2},p_{5},p_{6}}^{\left( 0\right) \left(
a,b,e,e\right) }\left( i\Omega _{n}\right) \Upsilon _{p_{5},p_{6}}^{\left(
e\right) }\left( \mathbf{q}\right) V\left( \mathbf{q}\right) \Upsilon
_{p_{7},p_{8}}^{\left( f\right) }\left( -\mathbf{q}\right) \chi
_{p_{7},p_{8},p_{3},p_{4}}^{\left( f,f,c,d\right) }\left( i\Omega
_{n}\right) \nonumber \\
&&-\frac{1}{\hbar}\frac{1}{{\cal V}}\sum_{e,f}\sum_{p_{5},p_{6},p_{7},p_{8}}%
\sum_{\mathbf{q}}\chi _{p_{1},p_{2},p_{5},p_{6}}^{\left( 0\right) \left(
a,b,e,f\right) }\left( i\Omega _{n}\right) \Upsilon _{p_{5},p_{8}}^{\left(
e\right) }\left( \mathbf{q}\right) V\left( \mathbf{q}\right) \Upsilon
_{p_{7},p_{6}}^{\left( f\right) }\left( -\mathbf{q}\right) \chi
_{p_{7},p_{8},p_{3},p_{4}}^{\left( f,e,c,d\right) }\left( i\Omega
_{n}\right) , \nonumber
\end{eqnarray}%
where $V({\bf q})=V\left( \mathbf{q}_{\bot },q_{z}\right) =e^{2}/ (\varepsilon_{\infty} \left( q_{\bot }^{2}+q_{z}^{2}\right)) $ and we have defined
\begin{equation}
\chi _{p_{1}, p_2, p_3, p_{4}}^{\left( 0\right) \left( a,b,c,d\right) }\left(
i\Omega _{n}\right) =\frac{1}{\beta \hbar }\sum_{\omega
_{n}}G_{p_{2},p_{3}}^{\left( b,c\right) }\left( i\omega_n+i\Omega_n\right)
G_{p_{4},p_{1}}^{\left( d,a\right) }\left( i\omega _{n}\right).
\label{grpa11}
\end{equation}%
Also, Eq. (\ref{grpa6}) contains the matrix element of the Fourier transform of the single-particle density operator,
\begin{equation}
  \label{eq:Upsi}
\Upsilon _{p_{1},p_{2}}^{\left( a\right) }\left( \mathbf{q}\right) =\int d%
\mathbf{u}\,e^{-i\mathbf{q}\cdot 
\mathbf{u}} \left[w_{a,p_{1}}^{\dag }\left(\mathbf{u}\right)\cdot w_{a,p_{2}}\left( \mathbf{u}\right)\right] .
\end{equation}
We now consider a magnetic field  $\mathbf{B}=\mathbf{\nabla }\times \mathbf{%
  A=}B\hat{\bf z}$.
In the Landau gauge, where $\mathbf{A}=\left(0,Bx,0\right)$, the eigenspinors $w_{a,p}\left( \mathbf{u}\right) $
are given by
\begin{equation}
  \label{eq:eigenf}
w_{a,p}\left( \mathbf{u}\right) = \frac{1}{\sqrt{L_{z}}}%
w_{a,n,k,X}\left( \mathbf{r}\right) e^{-ikz},
\end{equation}%
where $k$ is the momentum parallel to the magnetic field, $L_z$ is the sample length in the $z$ direction, $n$ is the Landau level index ($n=0$ for the chiral LL, $n>0$ for the conduction band LLs, $n<0$ for the valence band LLs), and $X$ is the guiding-center index.
The energy eigenvalues are degenerate in $X$
with degeneracy $N_{\varphi }=S/2\pi \ell_B
^{2}$, where $S=L_x L_y$ is the area of the system in the $xy$ plane and $\ell_B =\sqrt{%
  \hslash /eB}$ is the magnetic length.

In this basis, the field operator becomes
\begin{equation}\label{eq:basis}
\Psi _{a}\left( \mathbf{u}\right) = \frac{1}{\sqrt{L_{z}}} 
\sum_{n,X,k}w_{a,n,k,X}\left( \mathbf{r}\right) e^{-ikz}c_{a,n,k,X} = \frac{1}{\sqrt{L_{z}}} 
\sum_{n,X,k}
\begin{pmatrix}
u_{n,k,a} h_{|n|-1,X} (\mathbf{r}) \\ v_{n,k,a} h_{|n|,X} (\mathbf{r})
\end{pmatrix}
 e^{-ikz}c_{a,n,k,X},
\end{equation}
where $h_{|n|,X} = ((-i)^{|n|}/\sqrt{L_y}) \varphi_{|n|} (x-X) \exp(-iXy/\ell_B^2) $ and $\varphi_{|n|} (x)$ are the eigenfunctions of the 1D quantum harmonic oscillator. We also have  $h_{n < 0,X}(\mathbf{r}) =0$.
The coefficients $u_{n,k,a}$ and $v_{n,k,a}$ are given for example in Ref. [\onlinecite{Tchoumakov2016}].

For a Weyl semimetal with tilted cones, the second equality in Eq.~(\ref{eq:basis}) is valid only when the magnetic field is parallel to the tilt vector.
Likewise, for a model of Weyl semimetal with nonlinear terms in the energy dispersion (see the next Appendix), the second equality in Eq.~(\ref{eq:basis}) is valid only when the magnetic field is oriented along the axis of cylindrical symmetry.
For the rest of this Appendix, we will assume these simple configurations and use the second equality of Eq.~(\ref{eq:basis}). 
In the more general case with an arbitrary orientation of the magnetic field, the form of the noninteracting eigenspinors is more complicated and the analysis of the Coulomb interactions becomes significantly more cumbersome.
In the main text, we have included various figures for the noninteracting optical absorption, in which the magnetic field was not parallel to the tilt vectors.
In those figures, we have obtained the eigenstates of the non-interacting Hamiltonian numerically, although an analytical approach is also feasible.\cite{Tchoumakov2016} 


Substituting Eq.~(\ref{eq:eigenf}) in Eq.~(\ref{eq:Upsi}), the latter becomes
\begin{eqnarray}
\label{eq:upsilon0}
  \Upsilon _{n_{1},k_{1},X_{1};n_{2},k_{2},X_{2}}^{\left( a\right) }\left( 
\mathbf{q}\right)  &=&\frac{1}{L_{z}}\int d\mathbf{u}%
\,e^{ik_{1}z}e^{-i%
\mathbf{q}\cdot \mathbf{u}}\,e^{-ik_{2}z}\left[w_{a,n_{1},k_{1},X_{1}}^{\dag }\left( \mathbf{r}\right)\cdot w_{a,n_{2},k_{2},X_{2}}\left( \mathbf{%
r}\right)\right] \nonumber \\
&=&\delta _{k_{1},k_{2}+q_{z}}\int d\mathbf{r}\,e^{-i\mathbf{q}_{\bot }\cdot \mathbf{r}%
}\left[w_{a,n_{1},k_{1},X_{1}}^{\dag
}\left( \mathbf{r}\right)\cdot w_{a,n_{2},k_{2},X_{2}}\left( \mathbf{r}\right)\right]   \nonumber \\
&\equiv &\delta _{k_{1},k_{2}+q_{z}}\widetilde{\Upsilon }%
_{n_{1}k_{2}+q_{z},n_{2}k_{2}}^{\left( a\right) }\left( -\mathbf{q}_{\bot
}\right) e^{-\frac{i}{2}q_{x}\left( X_{1}+X_{2}\right) }\delta
_{X_{1},X_{2}+q_{y}\ell_B ^{2}} ,
\end{eqnarray}%
where
\begin{align}
\widetilde{\Upsilon }_{n_{1}k_{2}+q_{z},n_{2}k_{2}}^{\left( a\right) }\left( -\mathbf{q}_{\bot
}\right) &= u^*_{n_1, k_2 +q_z, a} u_{n_2, k_2, a} F_{|n_1|-1,|n_2|-1}(- \mathbf{q_\perp}) + v^*_{n_1, k_2 +q_z, a} v_{n_2, k_2, a} F_{|n_1|,|n_2|}(- \mathbf{q_\perp}) \label{eq:ups} \\
F_{n_1,n_2}(\mathbf{q_\perp}) &= \sqrt{\frac{\min(n_1,n_2)!}{\max(n_1,n_2)!}} \left(\frac{\pm q_y \ell_B +iq_x \ell_B}{\sqrt{2}} \right)^{|n_1-n_2|} L_{\min(n_1,n_2)}^{|n_1-n_2|}(q_\perp^2 \ell_B^2 /2) e^{-q_\perp^2 \ell_B^2/4}.
\label{eq:F}
\end{align}
The expression in Eq.~(\ref{eq:F}) is derived in Ref. [\onlinecite{Grad}].
The $+$($-$) sign therein holds for $n_1 > n_2$ ($n_1 < n_2$), and $L_n^a(x)$ are the generalized Laguerre polynomials.
The relations $F_{n_1,n_2} (\mathbf{q_\perp}=0) = \delta_{n_1,n_2}$ and $\tilde{\Upsilon}^{(a)}_{n_1 k, n_2 k}({\bf 0}) = \delta_{n_1, n_2}$ will be useful below. 

Similarly, the matrix elements of the current operator can be expressed as
\begin{eqnarray}
\Lambda _{n_{1},k_{1},X_{1};n_{2},k_{2},X_{2}}^{\left( a,\alpha \right)
}\left( \mathbf{q}\right)  &=&\frac{1}{L_{z}}\int d\mathbf{u}%
\,e^{ik_1 z}e^{-i%
\mathbf{q}\cdot \mathbf{u}}e^{-ik_2 z}\left[w_{a,n_{1},k_{1},X_{1}}^{\dag }\left( \mathbf{u}\right)\cdot j_{a,\alpha
}\cdot w_{a,n_{2},k_{2},X_{2}}\left( \mathbf{u}\right)\right]   \label{grpa9} \\
&=&\delta _{k_{1},k_{2}+q_{z}}\int d\mathbf{r}\,e^{-i\mathbf{q}_{\bot }\cdot \mathbf{r}%
} \left[w_{a,n_{1},k_{1},X_{1}}^{\dag
}\left( \mathbf{r}\right)\cdot j_{a,\alpha } \cdot w_{a,n_{2},k_{2},X_{2}}\left( \mathbf{r}\right)\right]   \nonumber \\
&\equiv &\delta _{k_{1},k_{2}+q_{z}}\widetilde{\Lambda }%
_{n_{1}k_{2}+q_{z},n_{2}k_{2}}^{\left( a,\alpha \right) }\left( -\mathbf{q}%
_{\bot }\right) e^{-\frac{i}{2}q_{x}\left( X_{1}+X_{2}\right) }\delta
_{X_{1},X_{2}+q_{y}\ell_B ^{2}}, \nonumber
\end{eqnarray}%
where $\widetilde{\Lambda}_{n_1 k_1, n_2 k_2}^{(a,\alpha)}$ can be computed in a way similar to Eq.~(\ref{eq:ups}).
Since $\widetilde{\Upsilon }_{n_{1} k_{1},n_{2} k_{2}}^{\left( a\right)
}\left( \mathbf{q}_{\bot }\right) $ and $\widetilde{\Lambda }%
_{n_{1} k_{1},n_{2} k_{2}}^{\left( a,\alpha \right) }\left( \mathbf{q}_{\bot
}\right) $ do not contain the guiding-center index, the expression in Eq. (%
\ref{grpa1}) can be simplified as
\begin{eqnarray}
  \label{eq:J2}
J_{a,\alpha }\left( \mathbf{q}\right)  &=&\sum_{n_{1},n_{2},k}\widetilde{%
\Lambda }_{n_{1}k+q_{z},n_{2}k}^{a,\alpha }\left( -\mathbf{q}_{\bot }\right)
\sum_{X_{1},X_{2}}e^{-\frac{i}{2}q_{x}\left( X_{1}+X_{2}\right) }\delta
_{X_{1},X_{2}+q_{y}\ell_B ^{2}}c_{a,n_{1},k+q_{z},X_{1}}^{\dag
}c_{a,n_{2},k,X_{2}} \\
&=&N_{\varphi }\sum_{n_{1},n_{2},k}\widetilde{\Lambda }%
_{n_1 k+q_{z},n_{2}k_{z}}^{a,\alpha }\left( -\mathbf{q}_{\bot }\right) \rho
_{n_{1}k+q_{z},n_{2}k}^{\left( a,a\right) }\left( \mathbf{q}_{\bot }\right) ,
\nonumber
\end{eqnarray}%
where we have introduced the operator
\begin{equation}
\rho_{n_{1}k+q_{z},n_{2}k}^{\left( a,b\right) }\left( \mathbf{q}_{\bot
}\right) =\frac{1}{N_{\varphi }}\sum_{X_{1},X_{2}}e^{-\frac{i}{2}q_{x}\left(
X_{1}+X_{2}\right) }\delta _{X_{1},X_{2}+q_{y}\ell_B
^{2}}c_{a,n_{1},k+q_{z},X_{1}}^{\dag }c_{b,n_{2},k,X_{2}} . \label{eq:rho}
\end{equation}%
This operator is related to the Fourier transform of the electronic density operator,
\begin{eqnarray}
n_{e}\left( \mathbf{q}\right)  &=&\sum_{a}\int d\mathbf{u}\,e^{-i\mathbf{q}_{\bot }\cdot \mathbf{r}%
}e^{-iq_{z}z}
\Psi
_{a}^{\dag }\left( \mathbf{u}\right) \Psi _{a}\left( \mathbf{u}\right)  \\
&=&N_{\varphi }\sum_{a}\sum_{n_{1},n_{2},k}\widetilde{\Upsilon }%
_{n_{1}k+q_{z},n_{2}k}^{a,\alpha }\left( -\mathbf{q}_{\bot }\right) \rho
_{n_{1}k+q_{z},n_{2}k}^{\left( a,a\right) }\left( \mathbf{q_\perp} \right) . \nonumber
\end{eqnarray}%
Replacing Eq.~(\ref{eq:J2}) in Eq.~(\ref{grpa1}), the imaginary-time current response function becomes
\begin{equation}
\chi _{J_{\alpha }J_{\beta }}\left( \mathbf{q},i\Omega _{n}\right) =\frac{%
N_{\varphi }}{{\cal V}\hbar}\sum_{a,b}\sum_{n_{1} \cdots n_{4}}%
\sum_{k_{1},k_{3}}\widetilde{\Lambda }_{n_1 k_{1}+q_{z}, n_2 k_{1}}^{\left(a,\alpha \right) }\left( -\mathbf{q}_{\bot }\right)
P_{n_{1}k_{1}+q_{z},n_{2}k_{1},n_{3}k_{3}-q_{z},n_{4}k_{3}}^{\left(a,a,b,b\right) }\left( \mathbf{q}_{\bot },\mathbf{q}_{\bot };i\Omega
_{n}\right) \widetilde{\Lambda }_{n_{3}k_{3}-q_{z},n_{4} k_{3}}^{\left(
b,\beta \right) }\left( \mathbf{q}_{\bot }\right) , \label{grpa20}
\end{equation}%
where
\begin{equation}
P_{n_{1}k_{1},n_{2}k_{2},n_{3}k_{3},n_{4}k_{4}}^{\left( a,a,b,b\right)
}\left( \mathbf{q}_{\bot },\mathbf{q}_{\bot }^{\prime };\tau \right)
=-N_{\varphi }\left\langle T_{\tau }\rho _{n_{1}k_{1},n_{2}k_{2}}^{\left(
a,a\right) }\left( \mathbf{q}_{\bot },\tau \right) \rho
_{n_{3}k_{3},n_{4}k_{4}}^{\left( b,b\right) }\left( -\mathbf{q}_{\bot
}^{\prime },0\right) \right\rangle.  \label{grpa10}
\end{equation}
Combining Eqs.~(\ref{grpa10}), (\ref{eq:rho}), (\ref{eq:upsilon0}) and (\ref{grpa6}), we arrive at 
\begin{eqnarray}
&&P_{n_{1}k_{1},n_{2}k_{2},n_{3}k_{3},n_{4}k_{4}}^{\left( a,a,b,b\right)
}\left( \mathbf{q}_{\bot },\mathbf{q}_{\bot }^{\prime };i\Omega _{n}\right) 
= P_{n_{1}k_{1},n_{2}k_{2},n_{3}k_{3},n_{4}k_{4}}^{\left( 0\right) \left(
a,a,b,b\right) }\left( \mathbf{q}_{\bot },\mathbf{q}_{\bot }^{\prime
};i\Omega _{n}\right)   \label{grpa12} \\
&&+\frac{N_{\varphi }}{\hbar {\cal V}}\sum_{c,d}\sum_{\mathbf{p}%
}\sum_{k_{1}^{\prime },k_{3}^{\prime }}\sum_{n_{1}^{\prime
} \cdots n_{4}^{\prime }}P_{n_{1}k_{1},n_{2}k_{2},n_{3}^{\prime }k_{3}^{\prime
},n_{4}^{\prime }k_{3}^{\prime }+p_{z}}^{\left( 0\right) \left(
a,a,c,c\right) }\left( \mathbf{q}_{\bot },\mathbf{p}_{\bot };i\Omega
_{n}\right) H_{n_{3}^{\prime }k_{3}^{\prime },n_{4}^{\prime }k_{3}^{\prime
}+p_{z},n_{1}^{\prime }k_{1}^{\prime },n_{2}^{\prime }k_{1}^{\prime
  }-p_{z}}^{\left( c,d\right) }\left( \mathbf{p}\right) \nonumber\\
&&\times  P_{n_{1}^{\prime
}k_{1}^{\prime },n_{2}^{\prime }k_{1}^{\prime
}-p_{z},n_{3}k_{3},n_{4}k_{4}}^{\left( d,d,b,b\right) }\left( \mathbf{p}%
_{\bot },\mathbf{q}_{\bot }^{\prime };i\Omega _{n}\right)   \nonumber \\
&&-\frac{1}{\hbar {\cal V}}\sum_{c,d}\sum_{\mathbf{p}%
}\sum_{k_{1}^{\prime },k_{3}^{\prime }}\sum_{n_{1}^{\prime
} \cdots n_{4}^{\prime }}P_{n_{1}k_{1},n_{2}k_{2},n_{3}^{\prime }k_{3}^{\prime
},n_{4}^{\prime }k_{1}^{\prime }}^{\left( 0\right) \left( a,a,c,d\right)
}\left( \mathbf{q}_{\bot },\mathbf{p}_{\bot };i\Omega _{n}\right)
X_{n_{3}^{\prime }k_{3}^{\prime },n_{2}^{\prime }k_{3}^{\prime
}-p_{z},n_{1}^{\prime }k_{1}^{\prime }-p_{z},n_{4}^{\prime }k_{1}^{\prime
}}^{\left( c,d\right) }\left( \mathbf{p}\right) \nonumber \\
&& \times P_{n_{1}^{\prime
}k_{1}^{\prime }-p_{z},n_{2}^{\prime }k_{3}^{\prime
}-p_{z},n_{3}k_{3},n_{4}k_{4}}^{\left( d,c,b,b\right) }\left( \mathbf{p}%
_{\bot },\mathbf{q}_{\bot }^{\prime };i\Omega _{n}\right), \nonumber
\end{eqnarray}%
where the Hartree and Fock interactions are defined as
\begin{eqnarray}
H_{n_{1} k_{1},n_{2}k_2,n_{3}k_{3},n_{4}k_4}^{\left(
c,d\right) }\left( \mathbf{p}\right)  &=&\widetilde{\Upsilon }%
_{n_{1}k_{1};n_{2}k_2}^{\left( c\right) }\left( \mathbf{p}_{\bot
}\right) V\left( \mathbf{p}\right) \widetilde{\Upsilon }%
_{n_{3}k_{3};n_{4}k_4}^{\left( d\right) }\left( -\mathbf{p}_{\bot
}\right) \label{vertex} \\
X_{n_{1}k_{1},n_{2}k_2,n_{3}k_{3},n_{4}k_{4}}^{\left(
c,d\right) }\left( \mathbf{p}\right)  &=&\sum_{\mathbf{t}_{\bot
}}e^{-i\left( \mathbf{t}_{\bot }\times \mathbf{p}_{\bot }\right) \cdot 
\hat{\bf z}\, \ell_B ^{2}}\widetilde{\Upsilon }%
_{n_{1}k_{1};n_{2}k_{2}}^{\left( c\right) }\left( \mathbf{t}_{\bot
}\right) V\left( \mathbf{t}_{\bot },p_{z}\right) \widetilde{\Upsilon }%
_{n_{3}k_{3};n_{4}k_{4}}^{\left( d\right) }\left( -\mathbf{t}_{\bot
}\right).  \nonumber
\end{eqnarray}
In Eq.~(\ref{vertex}), ${\bf t}_\perp$ is a two-dimensional momentum in the $xy$ plane (not to be confused with the tilt vector of a Weyl cone).
In the derivation of the Fock term, we also used the relation
\begin{equation}
c^\dagger_{a, n, k, X} c_{b, n', k', X'} = \sum_{{\bf p}_\perp} \rho_{n k,n'k'}^{\left( a,b\right) }\left( \mathbf{p}_{\perp}\right) e^{\frac{i}{2} p_x (X+X')} \delta_{X,X'+p_y \ell_B^2} . \label{eq:rev_rho}
\end{equation}
The Hartree-Fock two-particle Green's function $P_{n_{1}k_{1},n_{2}k_{2},n_{3}k_{3},n_{4}k_{4}}^{\left( 0\right) \left(a,a,b,b\right) }\left( \mathbf{q}_{\bot },\mathbf{q}_{\bot }^{\prime};i\Omega _{n}\right) $ appearing in Eq. (\ref{grpa12}) can be obtained by computing Eq. (\ref{grpa11}) or by solving the equation of motion of the Hartree-Fock Hamiltonian.
We will take the latter approach. 

The Hartree-Fock Hamiltonian $H_{HF}$ is obtained by performing the Hartree-Fock pairing in Eq. (\ref{grpa22}).
The outcome reads
\begin{eqnarray}
H_{HF}-\mu N &&= N_{\varphi }\sum_{a,n,k}\left[ E_{n,k}^{\left( a\right)
}-\mu \right] \rho _{nk,nk}^{\left( a,a\right) }\left( \mathbf{q}_{\bot
}=0\right)  \\
&&+\frac{N_{\varphi }^{2}}{{\cal V}}\sum_{\mathbf{q}}\sum_{a,b}%
\sum_{k_{1},k_{2}}%
\sum_{n_{1}\cdots n_{4}}H_{n_{1}k_{1},n_{4}k_{1}+q_{z},n_{2}k_{2},n_{3}k_{2}-q_{z}}^{\left( a,b\right) }\left( 
\mathbf{q}\right) \left\langle \rho _{n_{1}k_{1},n_{4}k_{1}+q_{z}}^{\left(
a,a\right) }\left( -\mathbf{q}_{\bot }\right) \right\rangle \rho
_{n_{2}k_{2},n_{3}k_{2}-q_{z}}^{\left( b,b\right) }\left( \mathbf{q}_{\bot
}\right)   \nonumber \\
&&-\frac{N_{\varphi }}{{\cal V}}\sum_{\mathbf{q}}\sum_{a,b}\sum_{k_{1},k_{2}}%
\sum_{n_{1}\cdots n_{4}}X_{n_{1}k_{1},n_{4}k_{1}+q_{z};n_{2}k_{2},n_{3}k_{2}-q_{z}}^{\left( a,b\right) }\left( 
\mathbf{q}\right) \left\langle \rho_{n_{1}k_{1},n_{3}k_{2}-q_{z}}^{\left(
a,b\right) }\left( -\mathbf{q}_{\bot }\right) \right\rangle \rho
_{n_{2},k_{2},n_{4}k_{1}+q_{z}}^{\left( b,a\right) }\left( \mathbf{q}_{\bot
}\right) , \nonumber
\end{eqnarray}%
where $E_{n,k}^{\left( a\right) }$ are the non-interacting energies.

We are now able find an equation for $P^{(0)}$ using the equation of motion $ \hbar \partial \rho(\tau)/ \partial \tau=[H_{HF}-\mu N,\rho(\tau)]  $.
A lenghty calculation yields
\begin{eqnarray}
&&\left[ i\Omega _{n}-\left( E_{n_{2}k_{2}}^{\left( b\right)
}-E_{n_{1}k_{1}}^{\left( a\right) }\right) /\hslash \right]
P_{n_{1}k_{1},n_{2}k_{2},n_{3}k_{3},n_{4}k_{4}}^{\left( 0\right) \left(
a,b,c,d\right) }\left( \mathbf{q}_{\bot },\mathbf{q}_{\bot }^{\prime
};i\Omega _{n}\right) =  \label{grpa13} \\
&&\left\langle \rho_{n_{1}k_{1},n_{4}k_{4}}^{\left( a,d\right) }\left( 
\mathbf{q}_{\bot }-\mathbf{q}_{\bot }^{\prime }\right) \right\rangle e^{i%
(\mathbf{q}_{\bot }\times \mathbf{q}_{\bot }^{\prime }) \cdot 
\hat{\bf z}\, \ell_B ^{2}/2}\delta_{b,c}\delta _{n_{2},n_{3}}\delta _{k_{2},k_{3}} -\left\langle \rho _{n_{3}k_{3},n_{2}k_{2}}^{\left( c,b\right) }\left( \mathbf{q}_{\bot }-\mathbf{q}_{\bot }^{\prime }\right) \right\rangle e^{-i (\mathbf{q}_{\bot }\times \mathbf{q}_{\bot }^{\prime }) \cdot \hat{\bf z} \ell_B ^{2}/2}\delta_{a,d}\delta _{n_{1},n_{4}}\delta _{k_{1},k_{4}}  \nonumber \\
&&-\frac{N_{\varphi }}{\hbar {\cal V}}\sum_{\mathbf{p}}\sum_{a^{\prime
}}\sum_{k_{1}^{\prime }}\sum_{n_{1}^{\prime }n_{2}^{\prime }n_{3}^{\prime
}} e^{-i (\mathbf{p}_{\bot }\times \mathbf{q}_{\bot }) \cdot 
\hat{\bf z}\, \ell_B^{2}/2} H_{n_{1}^{\prime }k_{1}^{\prime },n_{3}^{\prime }k_{1}^{\prime
}+p_{z},n_{2}^{\prime }k_{1}+p_{z},n_{1}k_{1}}^{\left( a^{\prime },a\right)
}\left( \mathbf{p}-\mathbf{q}_\perp\right) \left\langle \rho _{n_{1}^{\prime
}k_{1}^{\prime },n_{3}^{\prime }k_{1}^{\prime }+p_{z}}^{\left( a^{\prime
},a^{\prime }\right) }\left( \mathbf{q}_{\bot }-\mathbf{p}_{\bot }\right)
\right\rangle  \nonumber \\
&&\times P_{n_{2}^{\prime}k_{1}+p_{z},n_{2}k_{2},n_{3}k_{3},n_{4}k_{4}}^{\left( 0\right) \left(
a,b,c,d\right) }\left( \mathbf{p}_{\bot },\mathbf{q}_{\bot }^{\prime
};i\Omega _{n}\right)  \nonumber \\
&&+\frac{N_{\varphi }}{\hbar {\cal V}}\sum_{\mathbf{p}}\sum_{b^{\prime
}}\sum_{k_{1}^{\prime }}\sum_{n_{1}^{\prime } n_{2}^{\prime } n_{3}^{\prime
}} e^{i (\mathbf{p}_{\bot }\times \mathbf{q}_{\bot }) \cdot 
\hat{\bf z}\,\ell_B^{2}/2} H_{n_{1}^{\prime }k_{1}^{\prime },n_{2}^{\prime }k_{1}^{\prime
}+p_{z},n_{2}k_{2},n_{3}^{\prime }k_{2}-p_{z}}^{\left( b^{\prime },b\right)
}\left( \mathbf{p}-\mathbf{q}_\perp\right) \left\langle \rho _{n_{1}^{\prime
}k_{1}^{\prime },n_{2}^{\prime }k_{1}^{\prime }+p_{z}}^{\left( b^{\prime
}b^{\prime }\right) }\left( \mathbf{q}_{\bot }-\mathbf{p}_{\bot }\right)
\right\rangle  \nonumber \\
&&\times P_{n_{1}k_{1},n_{3}^{\prime}k_{2}-p_{z},n_{3}k_{3},n_{4}k_{4}}^{\left( 0\right) \left( a,b,c,d\right)
}\left( \mathbf{p}_{\bot },\mathbf{q}_{\bot }^{\prime };i\Omega _{n}\right) 
\nonumber \\
&&+\frac{1}{\hbar {\cal V}}\sum_{\mathbf{p}}\sum_{a^{\prime
}}\sum_{k_{2}^{\prime }}\sum_{n_{1}^{\prime }n_{2}^{\prime }n_{3}^{\prime
}} e^{-i (\mathbf{p}_{\bot }\times \mathbf{q}_{\bot } ) \cdot 
\hat{\bf z}\,\ell_B^{2}/2} X_{n_{1}^{\prime }k_{1}-p_{z},n_{1}k_{1},n_{2}^{\prime }k_{2}^{\prime
},n_{3}^{\prime }k_{2}^{\prime }-p_{z}}^{\left( a,a^{\prime }\right) }\left( 
\mathbf{p}-\mathbf{q}_\perp\right) \left\langle \rho_{n_{1}^{\prime
}k_{1}-p_{z},n_{3}^{\prime }k_{2}^{\prime }-p_{z}}^{\left( a,a^{\prime
}\right) }\left( \mathbf{q}_{\bot }-\mathbf{p}_{\bot }\right) \right\rangle 
\nonumber \\
&&\times P_{n_{2}^{\prime }k_{2}^{\prime
},n_{2}k_{2},n_{3}k_{3},n_{4}k_{4}}^{\left( 0\right) \left( a^{\prime
},b,c,d\right) }\left( \mathbf{p}_{\bot },\mathbf{q}_{\bot }^{\prime
};i\Omega _{n}\right)  \nonumber \\
&&-\frac{1}{\hbar {\cal V}}\sum_{\mathbf{p}}\sum_{b^{\prime
}}\sum_{k_{1}^{\prime }}\sum_{n_{1}^{\prime }n_{2}^{\prime }n_{3}^{\prime
}} e^{i (\mathbf{p}_{\bot }\times \mathbf{q}_{\bot })\cdot 
\hat{\bf z}\,\ell_B^{2}/2} X_{n_{1}^{\prime }k_{1}^{\prime }-p_{z},n_{2}^{\prime }k_{1}^{\prime
},n_{2}k_{2},n_{3}^{\prime }k_{2}-p_{z}}^{\left( b^{\prime },b\right)
}\left( \mathbf{p}-\mathbf{q}_\perp\right) \left\langle \rho _{n_{1}^{\prime
}k_{1}^{\prime }-p_{z},n_{3}^{\prime }k_{2}-p_{z}}^{\left( b^{\prime
},b\right) }\left( \mathbf{p}_{\bot }-\mathbf{q}_{\bot }\right) \right\rangle
\nonumber \\
&&\times P_{n_{1}k_{1},n_{2}^{\prime }k_{1}^{\prime
},n_{3}k_{3},n_{4}k_{4}}^{\left( 0\right) \left( a,b^{\prime },c,d\right)
}\left( \mathbf{p}_{\bot },\mathbf{q}_{\bot }^{\prime };i\Omega _{n}\right) .
\nonumber
\end{eqnarray}
Equations (\ref{grpa12}) and (\ref{grpa13}) are very general. They can be
used to study non-uniform states of the electron gas or states with Landau level mixing
and/or node coherence (i.e. states with nonzero averages such as $%
\left\langle \rho ^{\left( a,b\right) }\right\rangle $ with $a\neq b$)$.$ In
this paper, we are solely interested in uniform states with no coherence of
any sort. Hence,  we simplify the Hartree-Fock response with the condition
\begin{equation}\label{eq:liq}
\left\langle \rho _{n_1 k_{1},n_{2} k_{2}}^{\left( a,b \right) }\left( \mathbf{p}_{\bot }-\mathbf{q}_{\bot }\right) \right\rangle = \left\langle \rho _{n_1 k_1,n_1 k_1}^{\left( a,a \right) }\left( {\bf 0} \right) \right\rangle \delta_{k_1,k_2} \delta_{n_1, n_2} \delta_{a,b} \delta_{\mathbf{q_\perp, p_\perp}}.
\end{equation}
Using Eq.~(\ref{eq:liq}), the Hartree terms in Eq.~(\ref{grpa13}) involve $H$ functions evaluated at zero momentum; accordingly, they will be cancelled by the neutralizing positive
background.
In the last two terms of Eq.~(\ref{grpa13}), $X^{(a,a)}_{n_1' k_1-p_z, n_1 k_1, n_2' k_1, n_1' k_1-p_z} ({\bf 0},p_z) \propto \delta_{|n_2'|, |n_1|}$ and
$X^{(b,b)}_{n_1', k_2-p_z, n_2' k_2, n_2 k_2, n_1' k_2-p_z} ({\bf 0},p_z) \propto \delta_{|n_2'|, |n_2|}$.
These conditions can be demonstrated from Eq.~(\ref{vertex}).
In the preceding Kronecker deltas, the terms with $n_2'=-n_1$ and $n_2'=-n_2$ are subdominant compared to the terms with $n_2'=n_1$ and $n_2'=n_2$.
One reason for this is that $X$ has a logarithmic singularity at $p_z=0$ when (and only when) $n_2'=n_1$ and $n_2'=n_2$.
Hence, hereafter we keep only the latter terms.
Accordingly, we are left with
\begin{eqnarray}
&&\left[ i\Omega _{n}-\left( \widetilde{E}_{n_{2}k_{2}}^{\left( b\right) }-%
\widetilde{E}_{n_{1}k_{1}}^{\left( a\right) }\right) /\hbar \right]
P_{n_{1}k_{1},n_{2}k_{2},n_{3}k_{3},n_{4}k_{4}}^{\left( 0\right) \left(
a,b,c,d\right) }\left( \mathbf{q}_{\bot },\mathbf{q}_{\bot }^{\prime
};i\Omega _{n}\right) = \label{grpa17} \\
&&\left[ \left\langle \rho _{n_{1}k_{1},n_{1}k_{1}}^{\left( a,a\right)
}\left( 0\right) \right\rangle -\left\langle \rho
_{n_{2}k_{2},n_{2}k_{2}}^{\left( b,b\right) }\left( 0\right) \right\rangle %
\right] \delta _{a,d}\delta _{b,c}\delta _{n_{1},n_{4}}\delta
_{n_{2},n_{3}}\delta _{k_{2},k_{3}}\delta _{k_{1},k_{4}}\delta _{\mathbf{q}%
_{\bot },\mathbf{q}_{\bot }^{\prime }} , \nonumber
\end{eqnarray}%
where we have defined the self-energies in the renormalized single-particle energies $\widetilde{E}_{nk}^{\left( a\right) }=E_{nk}^{\left( a\right) }+\Sigma
_{nk}^{\left( a\right) }$ as 
\begin{equation}
\Sigma _{nk}^{\left( a\right) }=-\frac{1}{{\cal V}}\sum_{p_{z}}\sum_{n^{\prime
}}X_{n^{\prime }k-p_{z},nk,nk,n^{\prime }k-p_{z}}^{\left( a,a\right) }\left({\bf 0},
p_z\right) \left\langle \rho _{n^{\prime }k-p_{z},n^{\prime }k-p_{z}}^{\left(
a,a\right) }\left({\bf 0}\right) \right\rangle.
\end{equation}
In Eq. (\ref{grpa17}), the factor $\delta_{n_1, n_4} \delta_{n_2,n_3}$ reflects the fact that the Hartree-Fock Green's functions have been approximated to be diagonal in the noninteracting Landau level index.
This approximation is a consequence of the terms neglected in the paragraph following Eq. (\ref{eq:liq}).

Substituting Eq. (\ref{grpa17}) in Eq.~(\ref{grpa12}), the latter simplifies to 
\begin{eqnarray}
&&P_{n_{1}k_{1},n_{2}k_{2},n_{3}k_{3},n_{4}k_{4}}^{\left( a,a,b,b\right)
}\left( \mathbf{q}_{\bot },\mathbf{q}_{\bot }^{\prime };i\Omega _{n}\right) 
= P_{n_{1}k_{1},n_{2}k_{2},n_{2}k_{2},n_{1}k_{1}}^{\left( 0\right) \left(
a,a,a,a\right) }\left( \mathbf{q}_{\bot },\mathbf{q}_{\bot };i\Omega
_{n}\right) \delta _{a,b}\delta _{n_{1},n_{4}}\delta
_{n_{2},n_{3}}\delta _{k_{2},k_{3}}\delta _{k_{1},k_{4}}\delta _{\mathbf{q}%
_{\bot },\mathbf{q}_{\bot }^{\prime }}  \nonumber \\
&&+\frac{N_{\varphi }}{ \hbar {\cal V}}\sum_{c}\sum_{k_{1}^{\prime
}}\sum_{n_{1}^{\prime } n_{2}^{\prime
}}P_{n_{1}k_{1},n_{2}k_{2},n_{2}k_{2},n_{1}k_{1}}^{\left( 0\right) \left(
a,a,a,a\right) }\left( \mathbf{q}_{\bot },\mathbf{q}_{\bot };i\Omega
_{n}\right) H_{n_{2}k_{2},n_{1}k_{1},n_{1}^{\prime }k_{1}^{\prime
},n_{2}^{\prime }k_{1}^{\prime }-k_{1}+k_{2}}^{\left( a,c\right) }\left(
{\bf q}_\perp,k_{1}-k_{2}\right) \nonumber \\
&&\times P_{n_{1}^{\prime }k_{1}^{\prime },n_{2}^{\prime
}k_{1}^{\prime }-k_{1}+k_{2},n_{3}k_{3},n_{4}k_{4}}^{\left( c,c,b,b\right)
}\left( \mathbf{q}_{\bot },\mathbf{q}_{\bot }^{\prime };i\Omega _{n}\right) 
\nonumber \\
&&-\frac{1}{\hbar {\cal V}}\sum_{p_{z}}\sum_{n_{1}^{\prime
}n_{2}^{\prime }}P_{n_{1}k_{1},n_{2}k_{2},n_{2}k_{2},n_{1}k_{1}}^{\left(
0\right) \left( a,a,a,a\right) }\left( \mathbf{q}_{\bot },\mathbf{q}_{\bot
};i\Omega _{n}\right) X_{n_{2}k_{2},n_{2}^{\prime }k_{2}-p_{z},n_{1}^{\prime
}k_{1}-p_{z},n_{1}k_{1}}^{\left( a,a\right) }\left({\bf q}_\perp,p_{z}\right) \nonumber \\
&& \times P_{n_{1}^{\prime }k_{1}-p_{z},n_{2}^{\prime
}k_{2}-p_{z},n_{3}k_{3},n_{4}k_{4}}^{\left( a,a,b,b\right) }\left( \mathbf{q}%
_{\bot },\mathbf{q}_{\bot }^{\prime };i\Omega _{n}\right) . \nonumber
\end{eqnarray}
To compute the optical absorption, we can hereafter set ${\bf q}_\perp = {\bf q}'_\perp=0$. Therefore, we concentrate on the equation
\begin{eqnarray}
&&P_{n_{1}k_{2},n_{2}k_{2},n_{3}k_{4},n_{4}k_{4}}^{\left( a,a,b,b\right)
}\left( i\Omega _{n}\right) = P_{n_{1}k_{2},n_{2}k_{2},n_{2}k_{2},n_{1}k_{2}}^{\left( 0\right) \left(
a,a,a,a\right) }\left( i\Omega _{n}\right)  \delta _{a,b}\delta _{n_{1},n_{4}}\delta
  _{n_{2},n_{3}}\delta _{k_{2},k_{3}}\delta _{k_{1},k_{4}}
  \nonumber \\
&&+\frac{N_{\varphi }}{\hbar {\cal V}}\sum_{c}\sum_{k_{1}^{\prime
}}\sum_{n_{1}^{\prime }n_{2}^{\prime
}}P_{n_{1}k_{2},n_{2}k_{2},n_{2}k_{2},n_{1}k_{2}}^{\left( 0\right) \left(
a,a,a,a\right) }\left( i\Omega _{n}\right)
H_{n_{2}k_{2},n_{1}k_{2},n_{1}^{\prime }k_{1}^{\prime },n_{2}^{\prime
}k_{1}^{\prime }}^{\left( a,c\right) }\left({\bf 0},k_1-k_2\right) P_{n_{1}^{\prime
}k_{1}^{\prime },n_{2}^{\prime }k_{1}^{\prime
},n_{3}k_{4},n_{4}k_{4}}^{\left( c,c,b,b\right) }\left( i\Omega _{n}\right) 
\nonumber \\
&&-\frac{1}{\hbar {\cal V}}\sum_{p_{z}}\sum_{n_{1}^{\prime
}n_{2}^{\prime }}P_{n_{1}k_{2},n_{2}k_{2},n_{2}k_{2},n_{1}k_{2}}^{\left(
0\right) \left( a,a,a,a\right) }\left( i\Omega _{n}\right)
X_{n_{2}k_{2},n_{2}^{\prime }k_{2}-p_{z},n_{1}^{\prime
}k_{2}-p_{z},n_{1}k_{2}}^{\left( a,a\right) }\left({\bf 0},p_{z}\right)
P_{n_{1}^{\prime }k_{2}-p_{z},n_{2}^{\prime
}k_{2}-p_{z},n_{3} k_{4},n_{4} k_{4}}^{\left( a,a,b,b\right) }\left( i\Omega
_{n}\right) . \nonumber
\end{eqnarray}%
Hereafter we will be interested in the interband optical conductivity.
This involves considering  inter-Landau-level transitions for which $n_{1}\neq
n_{2}.$ The Hartree term is then zero because of the orthogonality of the spinors
in Eq. (\ref{vertex}). We are thus left with the excitonic corrections only: 
\begin{eqnarray}
&&P_{n_{1}k_{2},n_{2}k_{2},n_{3}k_{4},n_{4}k_{4}}^{\left( a,a,b,b\right)
}\left( i\Omega _{n}\right)   = P_{n_{1}k_{2},n_{2}k_{2},n_{2}k_{2},n_{1}k_{2}}^{\left( 0\right) \left(a,a,a,a\right) }\left( i\Omega _{n}\right) \delta _{a,b}\delta _{n_{1},n_{4}}\delta
  _{n_{2},n_{3}}\delta _{k_{2},k_{4}}
    \label{grpa18} \\
&&-\frac{1}{\hbar {\cal V}}\sum_{p_{z}}\sum_{n_{1}^{\prime
}n_{2}^{\prime }}P_{n_{1}k_{2},n_{2}k_{2},n_{2}k_{2},n_{1}k_{2}}^{\left(
0\right) \left( a,a,a,a\right) }\left( i\Omega _{n}\right)
X_{n_{2}k_{2},n_{2}^{\prime }k_{2}-p_{z},n_{1}^{\prime
}k_{2}-p_{z},n_{1}k_{2}}^{\left( a,a\right) }\left( {\bf 0},p_{z}\right)
P_{n_{1}^{\prime }k_{2}-p_{z},n_{2}^{\prime
}k_{2}-p_{z},n_{3}k_{4},n_{4}k_{4}}^{\left( a,a,b,b\right) }\left( i\Omega
_{n}\right) . \nonumber
\end{eqnarray}%
Next, we multiply both sides of Eq. (\ref{grpa18}) by $\left[ i\Omega _{n}-\left( \widetilde{E}%
_{n_{2}k_{2}}^{\left( a\right) }-\widetilde{E}_{n_{1}k_{2}}^{\left( a\right)
}\right) /\hslash \right] $ and use Eq. (%
\ref{grpa17}) to get
\begin{eqnarray}
  \label{eq:Pab}
&&\left[ \omega +i\delta -\left( \widetilde{E}_{n_{2}k_{2}}^{\left( a\right)
}-\widetilde{E}_{n_{1}k_{2}}^{\left( a\right) }\right) /\hbar \right]
P_{n_{1}k_{2},n_{2}k_{2},n_{3}k_{4},n_{4}k_{4}}^{\left( a,a,b,b\right)
}\left( \omega \right)\\
&&= \left[ \left\langle \rho _{n_{1}k_{2},n_{1}k_{2}}^{\left( a,a\right)
}\left( {\bf 0}\right) \right\rangle -\left\langle \rho
_{n_{2}k_{2},n_{2}k_{2}}^{\left( a,a\right) }\left( {\bf 0}\right) \right\rangle %
\right] \delta _{a,b}\delta _{n_{1},n_{4}}\delta _{n_{2},n_{3}}\delta
_{k_{2},k_{4}}  \nonumber \\
&&-\frac{1}{\hbar {\cal V}}\sum_{p_{z}}\sum_{n_{1}^{\prime} n_{2}^{\prime }}\left[ \left\langle \rho_{n_{1}k_{2},n_{1}k_{2}}^{\left(a,a\right) }\left( {\bf 0}\right) \right\rangle -\left\langle \rho_{n_{2}k_{2},n_{2}k_{2}}^{\left(a,a\right) }\left( {\bf 0}\right) \right\rangle %
\right]  X_{n_{2}k_{2},n_{2}^{\prime }k_{2}-p_{z},n_{1}^{\prime}k_{2}-p_{z},n_{1}k_{2}}^{\left( a,a\right) }\left( {\bf 0},p_{z}\right)
P_{n_{1}^{\prime }k_{2}-p_{z},n_{2}^{\prime}k_{2}-p_{z},k_{4}n_{3},k_{4}n_{4}}^{\left( a,a,b,b\right) }\left( \omega \right)\nonumber, 
\end{eqnarray}%
where we have carried out the analytical continuation.
From Eq.~(\ref{eq:Pab}), it follows (order by order in perturbation theory) that $P_{n_1 k_2, n_2 k_2, n_3 k_4, n_4 k_4}^{(a,a,b,b)} = P^{(a, a, a, a)}_{n_1 k_2, n_2 k_2, n_3 k_4, n_4 k_4} \delta_{a, b}$.
In addition, it can be shown from Eq. (\ref{vertex}) that, for $n_1\neq n_2$, $X^{(a,a)}_{n_2 k_2, n_2' k_2-p_z, n_1' k_2-p_z, n_1 k_2} ({\bf 0},p_z) \propto \delta_{|n_2'|, |n_2|} \delta_{|n_1'|, |n_1|}$.
The terms involving $n_2'=n_2$ and $n_1'=n_1$ are dominant in the small $p_z$ regime in which $X$ makes the largest (logarithmically diverging) contribution.
This is because $\tilde{\Upsilon}^{(a)}_{n_1 k, n_2 k}({\bf 0}) = \delta_{n_1, n_2}$.
Keeping only the terms $n_2'=n_2$ and $n_1'=n_1$ then leads to $P_{n_1 k_2, n_2 k_2, n_3 k_4, n_4 k_4}^{(a,a,a,a)} = P^{(a, a, a, a)}_{n_1 k_2, n_2 k_2, n_2 k_4, n_1 k_4} \delta_{n_2, n_3} \delta_{n_1,n_4}$.
In sum, Eq. (\ref{eq:Pab}) gets simplified to
\begin{align}
  \label{eq:Pab2}
&\left[ \omega +i\delta -\left( \widetilde{E}_{n_{2}k_{2}}^{\left( a\right)
}-\widetilde{E}_{n_{1}k_{2}}^{\left( a\right) }\right) /\hbar \right]
P_{n_{1}k_{2},n_{2}k_{2},n_{2}k_{4},n_{1}k_{4}}^{\left( a,a,a,a\right)
}\left( \omega \right)\\
&= \left[ \left\langle \rho _{n_{1}k_{2},n_{1}k_{2}}^{\left( a,a\right)
}\left( {\bf 0}\right) \right\rangle -\left\langle \rho
_{n_{2}k_{2},n_{2}k_{2}}^{\left( a,a\right) }\left( {\bf 0}\right) \right\rangle %
\right] \delta_{k_{2},k_{4}}  \nonumber \\
&-\frac{1}{\hbar {\cal V}}\sum_{p_{z}}\left[ \left\langle \rho_{n_{1}k_{2},n_{1}k_{2}}^{\left(a,a\right) }\left( {\bf 0}\right) \right\rangle -\left\langle \rho_{n_{2}k_{2},n_{2}k_{2}}^{\left(a,a\right) }\left( {\bf 0}\right) \right\rangle %
\right]  X_{n_{2}k_{2},n_{2} k_{2}-p_{z},n_{1} k_{2}-p_{z},n_{1}k_{2}}^{\left( a,a\right) }\left( {\bf 0},p_{z}\right)
P_{n_{1} k_{2}-p_{z},n_{2} k_{2}-p_{z},k_{4}n_{2},k_{4}n_{1}}^{\left( a,a,a,a\right) }\left( \omega \right)\nonumber.
\end{align}
For each value of $\omega$, $n_1$, $n_2$ and $k_4$, Eq.~(\ref{eq:Pab2}) can be recasted in the form of a matrix inversion problem.
The dimension of this matrix is given by the number of momenta $k_2$ for which we calculate $P$.
We discretize\cite{macDonald2017} the momentum parallel to the magnetic field in 200 points (increasing this number does not change the results appreciably), and incorporate an ultraviolet cutoff.
We measure energies and momenta in units of $\hbar v_F/l_B$ and $l_B^{-1}$, respectively.
The cutoff in $k$ implies that the range of the dimensionless $k l_B$ will depend on the value of the magnetic field.  
For example, when $B = 20 \, \text{T} $ and $40 \, \text{T}$, we take respectively $k \ell_B \in [ -5, 5 ] $ and $ k l_B \in [-3.5, 3.5]$.
Moreover, to render the problem tractable, we work on a truncated Hilbert space that includes only the chiral Landau level and the first three nonchiral Landau levels in both the conduction and valence bands.
This truncation is justified in the strong-field and low-frequency regimes that we are interested in.

Using Eq. (\ref{grpa20}), the retarded current response for each node $a$ is then given by%
\begin{equation}
\chi _{J_{\alpha }J_{\beta }}^{R}\left( \mathbf{q}=0,\omega \right) =\frac{%
N_{\varphi }}{{\cal V}\hbar }\sum_{n_{1},n_{2}}^{\ast }\sum_{k_{1},k_{2}}%
\widetilde{\Lambda }_{k_{1}n_{1},k_{1}n_{2}}^{\left( a,\alpha \right)
}\left( {\bf 0}\right) P_{n_{1}k_{1},n_{2}k_{1},n_{2}k_{2},n_{1}k_{2}}^{\left(
a,a,a,a\right) }\left( \omega \right) \widetilde{\Lambda }%
_{n_{2}k_{2},n_{1},k_{2}}^{\left( a,\beta \right) }\left( {\bf 0}\right),
\label{grpa21}
\end{equation}%
where the star over the Landau level summation is a reminder that we keep only a limited set of levels.
Plugging the solution of Eq.~(\ref{eq:Pab2}) in Eq.~(\ref{grpa21}), we get the interband magneto-optical conductivity.

\section{Nonlinear model} \label{App_nonlin}

In the main text, we have focused on a minimal model for Weyl semimetals with tilted cones.
However, we have emphasized therein that the crucial element to obtain a complete valley polarization is the breaking of the antiunitary symmetry $\Theta$ defined in Sec.~\ref{sec:mod}.
This symmetry can be broken even for untilted Weyl cones, provided that we include nonlinear terms in the energy dispersion.

The objective of this appendix is to demonstrate that complete valley polarization and the splitting of interband transitions involving the chiral Landau level do occur in a particle-hole-symmetric (tiltless) energy dispersion with nonlinear terms.

To that end,  we replace Eq.~(\ref{eq:model2}) by \cite{Nagaosa2016, Bertrand2017} 
\begin{align}\label{eq:model3}
d_{1,0}(\mathbf{k}) &= 0 \nonumber \\
d_{1,x}(\mathbf{k}) &= \hbar v_F k_x (1+\alpha k_z) \nonumber \\
d_{1,y}(\mathbf{k}) &= \hbar v_F k_y (1+\alpha k_z) \nonumber \\ 
d_{1,z}(\mathbf{k}) &= \hbar v_z k_z + \beta (k_x^2 + k_y^2 -2 k_z^2)
\end{align}
for the $\tau=1$ Weyl node, where $v_z\neq v_F$ indicates a possible anisotropy in the Dirac velocity (which turns out to be unimportant for our purposes) and $\alpha$ and $\beta$ are nonlinear parameters. 
The Hamiltonians for the remaining Weyl nodes can be obtained through the application of mirror and time-reversal operations:
\begin{align}\label{eq:symnnlin}
&\tau=1 \rightarrow 2   :  (v_z,\alpha, \beta) \rightarrow (-v_z,-\alpha,\beta)\nonumber \\
&\tau=1 \rightarrow 3   :  (v_z,\alpha,\beta)  \rightarrow (v_z,-\alpha,-\beta)\nonumber \\
&\tau=1 \rightarrow 4   :  (v_z,\alpha,\beta) \rightarrow (-v_z,\alpha,-\beta).
\end{align}

Next, we consider the effect of a magnetic field oriented along the $z$ direction.
This direction is particularly simple in that it preserves the cylindrical symmetry of the model, thereby allowing for an analytical solution of the electronic structure.
Other directions of the magnetic field can be similarly considered, with a numerical solution of the noninteracting electronic structure.
In the gauge $\mathbf{A}= (0,Bx,0)$, the coefficients of the eigenspinors of the noninteracting Hamiltonian (see Eq. (\ref{eq:basis})) near the node $\tau$ are
\begin{align}
u_{n k_z \tau} &= -i \, \text{sgn}(n) \sqrt{\frac{1}{2}\left( 1 + \text{sgn}(n) \frac{\hbar v_z k_z + \beta \left( 2|n| / \ell_B^2 - 2 k_z^2 \right)}{E^{(\tau)}_{n k_z} + \beta/ \ell_B^2} \right) } \nonumber \\
v_{n k_z \tau} &= \sqrt{\frac{1}{2}\left( 1 - \text{sgn}(n) \frac{\hbar v_z k_z + \beta \left( 2|n| / \ell_B^2 - 2 k_z^2 \right)}{E^{(\tau)}_{n k_z} + \beta/ \ell_B^2} \right) } , 
\end{align}
where $n$ is the Landau level index.
For the chiral Landau level ($n=0$), $u_{n \tau k_z}=0$ and $v_{n \tau k_z}=1$.
The eigenenergies for $\tau=1$ are given by

\begin{align}
E_{0 k_z}^{(1)} &= - \hbar v_z k_z - \beta (\ell_B^{-2} - 2 k_z^2) \label{eq:E_0_nnlin} \\
E_{n k_z}^{(1)} &= -\frac{\beta}{\ell_B^2} + \text{sgn}(n) \sqrt{2 |n| \frac{(\hbar v_F )^2}{\ell_B^2}(1 +\alpha k_z)^2+ \left[\hbar v_z k_z + \beta \left( \frac{2|n|}{\ell_B^2} - 2 k_z^2 \right) \right]^2 } .\label{eq:E_n_nnlin}
\end{align}

\begin{figure}[t]
\includegraphics[width=0.48\textwidth]{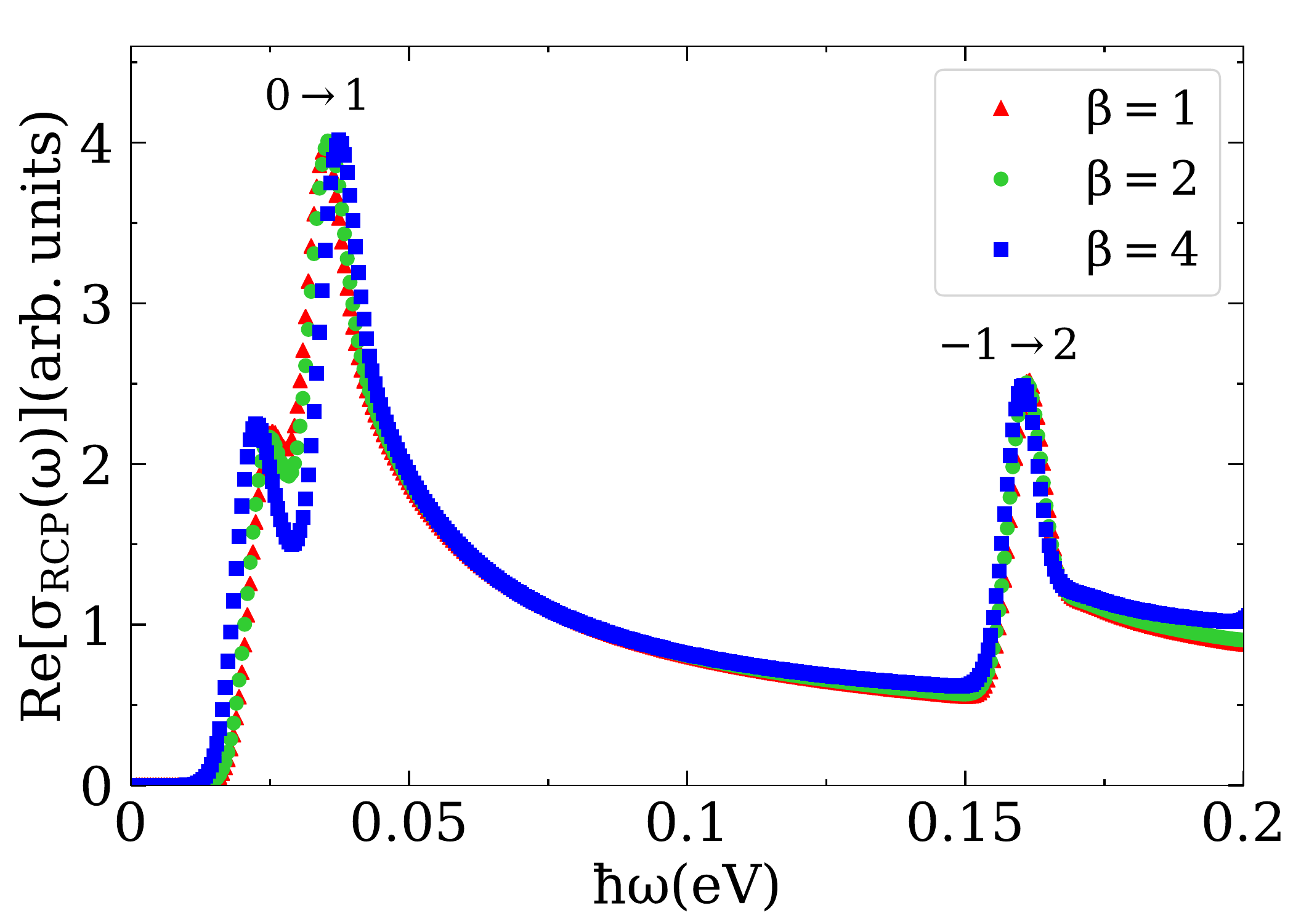}
\caption{Total optical absorption in a four-node WSM model, with ${\bf B}||{\bf q}||\hat{\bf z}$ and a RCP light. This figure is similar to Fig. \ref{fig:opttilt} of the main text, except that the tilt is replaced by nonlinear terms in the energy dispersion around the Weyl nodes.
  A splitting of the $0\to 1$ inter Landau level transition is evident.
  The parameters are $\alpha=0.08\ell_B$, $B=40 \text{T}$, $\epsilon_\infty=20$, $\mu=\hbar v_F / \ell_B$ and $\eta = 3 \text{meV}$.
  The red triangles, green dots and blue squares are for $\beta =1, \, 2$ and $4 \, \text{eV \AA}^2$ respectively. 
  These values of $\beta$ are of the same order of magnitude as the coefficients of the quadratic momentum terms in the energy dispersion of Dirac semimetals.\cite{Zwang2012,Zwang2013}
  }
\label{fig:optbeta}
\end{figure}

Starting from this model, we use the formalism from App. \ref{App_GRPA} in order to compute the optical conductivity.
Figure \ref{fig:optbeta} displays the result for the optical absorption, where a splitting of the $0 \to 1$ inter Landau level transition is apparent.
This splitting, induced by the nonlinear terms in the energy dispersion (i.e., by $\alpha$ and $\beta$),  is indicative of the complete valley polarization.
Thus, we get the same phenomenology as the tilted Weyl cone model with linear dispersion.

\end{widetext}

\bibliographystyle{apsrev4-1}
\bibliography{mybib}

\end{document}